\documentclass[
reprint,
superscriptaddress,
 amsmath,amssymb,
 aps,
]{revtex4-1}

\usepackage{graphicx}
\usepackage{dcolumn}
\usepackage{bm}
\usepackage{subcaption}
\usepackage[font=small,labelfont=bf,justification=raggedright, singlelinecheck=false]{caption}
\usepackage{natbib}
\usepackage[title]{appendix}

\usepackage{color}

\usepackage{url}

\begin{document}

\preprint{APS/123-QED}

\title{Velocity Distribution of a Homogeneously Cooling Granular Gas}
\author{Peidong Yu}
 \email{peidong.yu@dlr.de}
 \affiliation{Institut f\"ur Materialphysik im Weltraum, Deutsches  Zentrum f\"ur Luft- und Raumfahrt, 51170 K\"oln, Germany}
 \affiliation{Institut f\"ur Theoretische Physik, Universit\"at zu K\"oln, Z\"ulpicher Strasse 77, 50937 K\"oln, Germany}

\author{Matthias Schr\"oter}
 \email{schroeter@science-consulting.info}
 \affiliation{Institut f\"ur Materialphysik im Weltraum, Deutsches  Zentrum f\"ur Luft- und Raumfahrt, 51170 K\"oln, Germany}
 \affiliation{Hypatia - Science Consulting, 37081 G\"ottingen, Germany}

\author{Matthias Sperl}
 \email{matthias.sperl@dlr.de}
 \affiliation{Institut f\"ur Materialphysik im Weltraum, Deutsches  Zentrum f\"ur Luft- und Raumfahrt, 51170 K\"oln, Germany}
 \affiliation{Institut f\"ur Theoretische Physik, Universit\"at zu K\"oln, Z\"ulpicher Strasse 77, 50937 K\"oln, Germany}

\date{\today}

\begin{abstract}
In contrast to molecular gases, granular gases are characterized by inelastic collisions 
and require therefore permanent driving to maintain a constant kinetic energy.  
The kinetic theory of granular gases describes how the average velocity of the particles decreases
after the driving is shut off. Moreover it predicts that the rescaled particle velocity 
distribution will approach a stationary state with overpopulated high-velocity tails as compared to the Maxwell-Boltzmann distribution. While this fundamental theoretical result was reproduced by numerical simulations, an experimental confirmation is still missing. Using a microgravity experiment which allows the spatially homogeneous excitation of spheres
via magnetic fields, we confirm the theoretically predicted exponential decay of the tails of the velocity distribution.
\end{abstract}

\pacs{Valid PACS appear here}

\maketitle

\textit{Introduction}.--
Granular gases consist of macroscopic particles, i.e.~their diameter  $d$ is micrometers or larger. From this property follow two important differences between granular and molecular gases. First, all collisions between granular particles are \textit{inelastic}: a part of the mechanical energy of the relative motion of the particles is converted to heat \cite{Brilliantov2003}.
Secondly, the energy scales involved in granular dynamics, such as the energy needed to lift a grain its own diameter, are 10 to 20 orders of magnitude larger than the thermal energy of the system. In consequence, maintaining a dynamic state of a granular gas does require the constant injection of energy.

However, these differences to molecular gases also signify the chance to "reinvent statistical mechanics in a new context" \cite{kadanoff:99}. Properties of granular gases not known from their molecular counterpart include: non-Fourier heat flow \cite{soto:99,dufty:07,candela:07}, correlations \cite{brilliantov:07,kranz:09,gayen:11}, breaking of time-reversal symmetry \cite{poeschel:02,shaw:07}, segregation \cite{hsiau:96,jenkins:02,schroeter:06,garzo:06,garzo:11,brey:11,garzo:19,serero:08}, non-equipartition \cite{Goldshtein1995,feitosa:02,galvin:05,brilliantov:07,serero:08,harth:13}, and clustering \cite{goldhirsch:93,Goldhirsch1993_MDStudy,kudrolli:97,Falcon1999,opsomer:11,noirhomme:18,mitrano:12,Maass2008,hummel:16,opsomer:17}. Sometimes granular gases unexpectedly behave even like equilibrium systems \cite{nichol:12}. Motivated by these phenomena, theoretical physicist and mathematicians have started at the turn of the century to extend the kinetic theory from non-dissipative molecular systems \cite{Chapman1960} to dissipative particle systems;
for recent summaries at the textbook level see \cite{Brilliantov2003,garzo:19}.

The theoretical analysis of granular gases focuses on two  stationary states:
First the \textit{homogeneously driven gas} where energy is injected in a spatially homogeneous
way into the system in order to compensate the energy loss due to inelastic collisions. 
Kinetic theory predicts \cite{vanNoije1998} that the distribution of individual particle velocities $P(v)$ develops overpopulated tails when compared to the Maxwell-Boltzmann distribution:
$P(v) \sim \exp(- k v^{3/2} )$ for $v>\langle v\rangle/\epsilon$ where $\epsilon$ is the coefficient of restitution and  $\langle v\rangle$ is the average velocity of all particles. Second, the \textit{homogeneous cooling state} (HCS) where the system is not disturbed by external forces \cite{esipov:97,vanNoije1998,puglisi:99,goldhirsch:03,poeschel:06,poeschel:07}. This results in a monotonous decrease of $\langle v\rangle$ with time $t$ according to the so-called Haff's law \cite{Haff1983} :

 \begin{equation}
 \langle v(t)\rangle = v_0/ (1+t/\tau)^\gamma
 \label{eq:haff}
 \end{equation}
 $\tau$ is a material and density dependent characteristic time scale. If the coefficient of restitution $\epsilon$ of the particles can be considered to be velocity independent,  $\gamma$ equals one. If the velocity-dependence of $\epsilon$ for viscoelastic particles has to be taken into account, $\gamma$ becomes 5/6.
 
The reason that the HCS is qualified as a stationary state, even though $\langle v\rangle$ is a function of time,  is that the distribution of the rescaled particle velocities $c=v/v_T$, becomes stationary. 
$v_T$ is the velocity derived from granular temperature $T$ of the particles: $v_T = \sqrt{2T/m}$ where $m$ is the particle mass. 
As discussed in appendix \ref{APP:2}, this thermal velocity can be computed as
$v_{T} = 2 / \sqrt{\pi} \cdot \langle v \rangle $.
Kinetic theory \cite{esipov:97,vanNoije1998}  predicts that the tail of the velocity distribution  $P(c)$  becomes exponential, i.e.~$P(c) \sim \exp(-k c)$. The HCS ends with the onset of particle clustering which destroys its homogeneity \cite{hummel:16}.

Experimental confirmation of these kinetic theory results is difficult for two reasons. First, creating and maintaining a granular gas requires constant energy injection. Typically, this is done by vibrating the boundaries of the container, which however creates spatial and temporal inhomogeneities within the granular gas \cite{kudrolli:97,Falcon1999,bougie:02,meerson:04,eshuis:07,opsomer:11,noirhomme:18,Sack2013}, thereby invalidating one of the preconditions of the kinetic theory approaches. A spatially homogeneous driving can be obtained by either restricting the experiment to a horizontally aligned two-dimensional system which is shaken vertically \cite{olafsen:98,baxter:03,tatsumi:09,reis:06,deseigne:10}, or by using magnetic \cite{Maass2008,schmick:08,Falcon2013,Yu2019,Masato2019} or electrostatic \cite{aranson:02} forces to drive a three-dimensional system in the bulk.

Second, \textit{on earth} the gravitational field will always collapse the granular gas into a granular solid where collisions are replaced by enduring contacts.  This effect can again be counteracted by limiting the granular gas to a horizontally aligned two-dimensional system. This comes however at the price of a significantly larger number of particle-boundary than particle-particle collisions.

Force-free and three-dimensional granular gases can therefore only be realized by either levitating the particles in a magnetic field \cite{Maass2008} of by performing the experiments in a real 
microgravity environment, as e.g.~realized in parabolic flights \cite{Falcon2006,leconte:06,tatsumi:09,Sack2013,grasselli:15,noirhomme:18}, drop towers \cite{heiselmann:10,born:17}, sounding rockets \cite{Falcon1999,harth:13,harth:18}, or satellites \cite{Hou2008}.

In consequence, experimental confirmations of the main results of the kinetic theory of granular gases have been scarce. A number of groups tried to measure the the velocity distribution of the homogeneously driven gas using boundary-driven, two-dimensional \cite{olafsen:98,kudrolli:00,rouyer:00,baxter:03,blair:03,tatsumi:09} and three-dimensional \cite{losert:99,huan:04} systems. They reported high energy tails proportional to $\exp(- k v^{\alpha} )$ with values of $\alpha$ in the range 0.8 to 2. Especially for two-dimensional systems it has been shown that $\alpha$ is controlled by the friction coefficient of particle-sidewall collisions \cite{vanZon:04}. The only confirmations of $\alpha = 1.5$ were obtained in an electro-statically driven three-dimensional system \cite{aranson:02}, and in a two-dimensional, horizontal system containing rotationally-driven disks \cite{scholz:17}.

For the homogeneous cooling state, the validity of Eq. (\ref{eq:haff}) has been shown both in numerical simulations \cite{luding:98,kanzaki:10,bodrova:15,hummel:16} and experiments \cite{Maass2008,heiselmann:10,harth:18}. For the exponents $\alpha$  characterizing the tail of the velocity distribution values in the range 0.6 to 1.5 have been reported for boundary-driven, two-dimensional systems \cite{tatsumi:09}. The theoretically predicted exponential tail, i.e.~$\alpha =1$, has up to now only been seen in numerical simulations \cite{brey:99,brey:99-2}; an  experimental confirmation is missing.

In this letter, we describe a granular gas experiment which is both force-free, due to being performed in microgravity, and homogeneously heated via magnetic excitation. Our results confirm Haff's cooling law and show clear evidence of an exponential tail of the particle velocity distribution, thus directly confirming the validity of the kinetic theory of granular gases in the homogeneous cooling state.

\begin{figure}[t]
\centering
\includegraphics[width=0.35\textwidth]{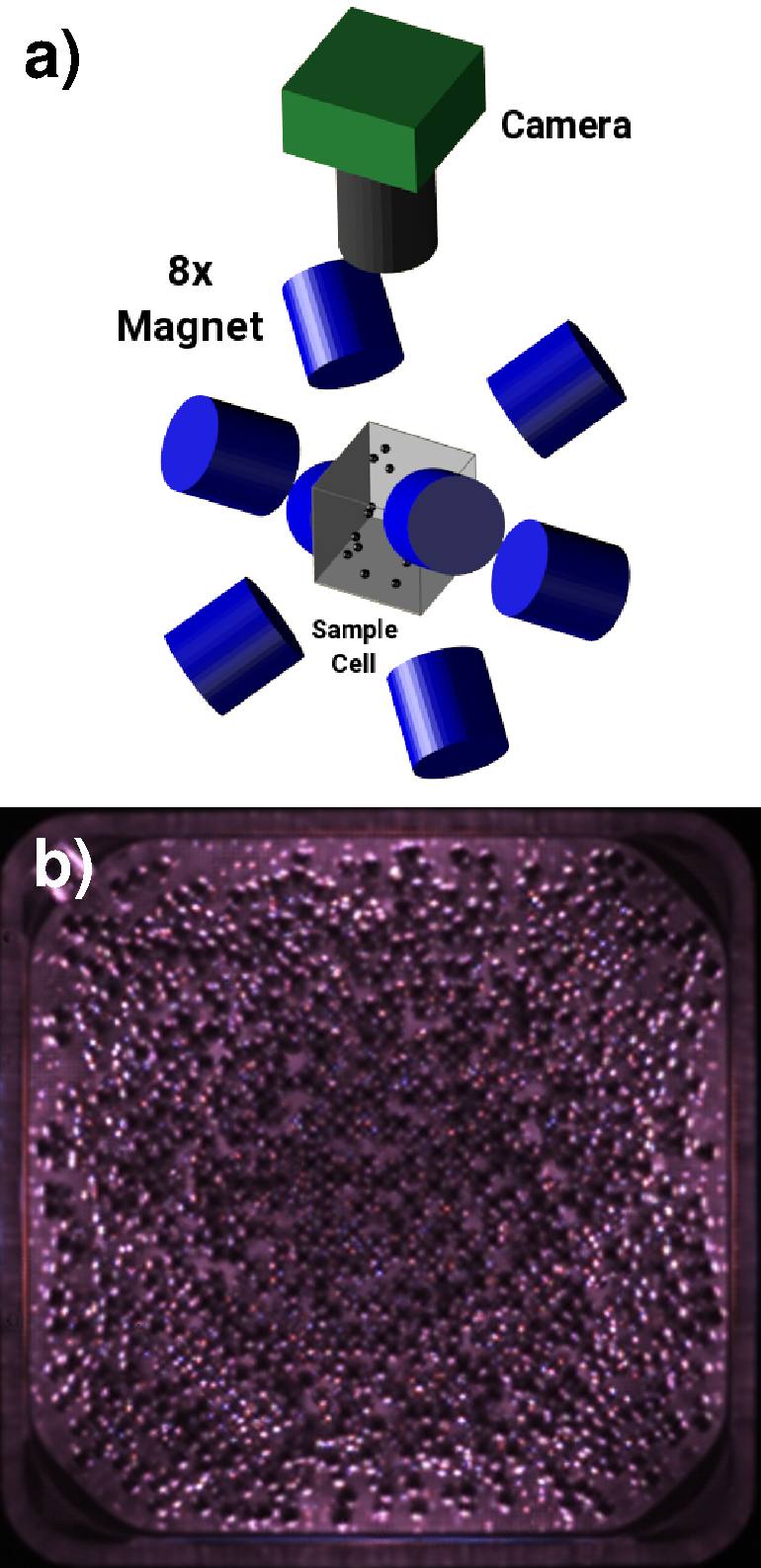}
\caption{Experimental setup. a) The diagonals of a PMMA cube 
containing 2796 ferromagnetic spheres are aligned with  four pairs of electromagnets which are used to inject energy into the granular gas. The time evolution of the granular gas is then observed with a camera at 165 frames per second.  Panel b) shows a typical example image.}
\label{fig:setup}
\end{figure}

\begin{figure*}[ht]
\centering
\includegraphics[width=0.9 \textwidth]{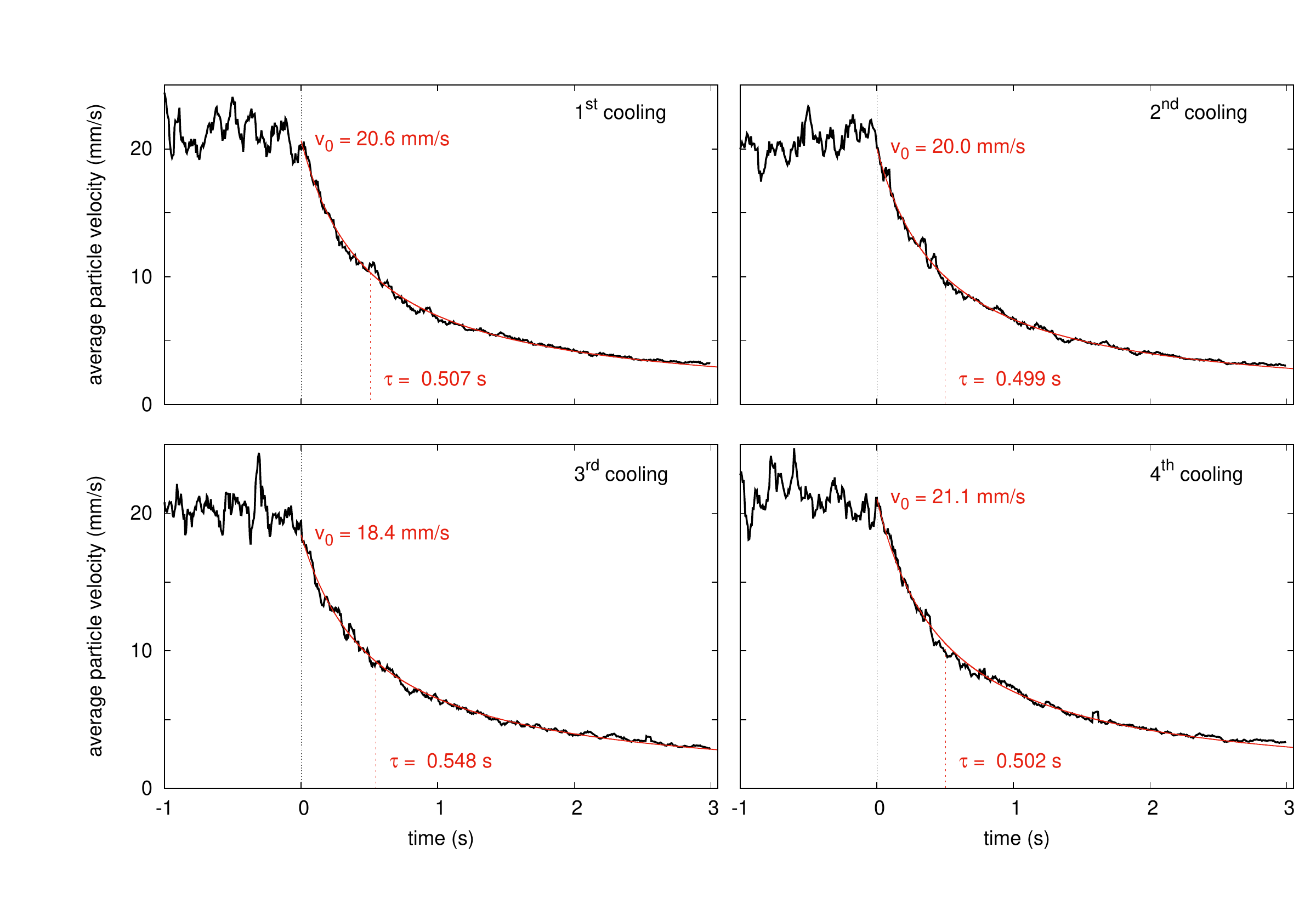}
\caption{All four experiments are well described by Haff's cooling law. Each panel displays first the last second of the magnetic driving with fluctuations in the average velocity. At $t=0$ the magnetic driving is switched off and the subsequent evolution of the average velocity is fit with Haff's law (Eq.~\ref{eq:haff}) assuming a constant coefficient of restitution (red line). Each fit provides two parameters: $v_0$, the average particle velocity at $t=0$, and $\tau$, the time by which the velocity has decreased to $v_0/2$.}
\label{fig:4xHaff}
\end{figure*}

\textit{Experimental Setup.--} The experiment was performed in a sounding rocket of the MAPHEUS campaign of German Aerospace Center (DLR) which provided 375 s of microgravity with a remnant gravity on the order of $10^{-5}$ times the earth acceleration \cite{Yu2019}. The granular gas consists of 2796 spheres with a diameter $d=1.6\pm0.02$ mm. The particles are contained in a cubic container made of PMMA acrylic glass with an inner side length of 50 mm. Taking into account the slanted corners of the cell,
the average volume fraction of the gas is $\phi=0.05$ and the number density $n=0.0234$ mm\textsuperscript{-3}. The interior of the container is connected to the outside atmosphere, resulting in an air pressure $<$ 0.01 Pa during the experiment. 
The resulting air drag reduces the particle velocity by 0.01\% per second \cite{Yu2019}.

The spheres are made of ferro-magnetic MuMetall (Sekels GmbH). Energy is injected into all granular particles in the bulk by pairs of magnets which are aligned along the diagonals of the cube, cf.~Fig.~\ref{fig:setup}. The excitation protocol cycles through the four pairs, switching each one on for 20 ms and then pausing for 80 ms. The low coercivity of the particles ensures that the long-range interactions between the particles are negligible during cooling \cite{Yu2019}, while during heating, such long-range interactions actually contribute to a more isotropic excitation \cite{Masato2019}. As shown in numerical simulations \cite{Masato2019} and experiments \cite{Yu2019}, this type of excitation results in a homogeneously heated granular gas. It can also be seen in the movie of the experiment which is part of the supplementary material. 

The granular dynamics inside the cell is observed with light field camera Raytrix R5 at 165 frames/s and a magnification of 70 $\mu$m/pixel. Due to its reduced precision, the depth information provided by the light field camera is not used in our analysis. Appendix \ref{APP:1}  provides further information on the two-dimensional image analysis.

\textit{Freely cooling granular gas}.--During the flight of the sounding rocket four individual experiments were performed. Each experiment consisted of a first phase of at least 40 s duration during which the granular gas was heated by switching the magnetic fields on and off. Then the excitation was stopped and a cooling period of at least 40 s duration followed. However, as shown in appendix \ref{APP:3}, after about 4 seconds the remnant gravitation starts to induce spatial inhomogeneities in the sample cell. We therefore limit the subsequent analysis to the first 3 seconds of the cooling phase; these were sufficient to reach a stationary state as shown in appendix \ref{APP:6}.

Fig.~\ref{fig:4xHaff} presents the evolution of the average particle velocity in the last second of the heating and the first three seconds of the cooling phase. Two features of Fig.~\ref{fig:4xHaff} are important to point out. First, Haff's law (Eq. (\ref{eq:haff}), displayed in red) provides a good fit to all four experiments. We have tested both exponents $\gamma$ = 1 and 5/6; the differences are within our experimental uncertainty. This signifies that the velocity-dependence of $\epsilon$ plays only a minor role in the velocity range analyzed here. 

This is to be expected because for $\epsilon (g) = 1 - \zeta g^{1/5} + h.o.t.$ and within the studied range of relative collision velocities $g$, the change of epsilon is small  ($\zeta$ contains material parameters) \cite{Brilliantov2003}. We assume in our subsequent analysis a constant $\epsilon=0.66$, which is measured from lab calibrations (see appendix \ref{APP:7} for details).

Second, the reproducibility is high between the four experiments. The parameters $v_0$ and $\tau$ of the Haff fits agree within $\pm 8\%$, and  $v_0\cdot\tau$, which is determined by system properties alone, agrees within $\pm 3\%$. Additionally, the particle velocities averaged over the last two seconds of the heating phase differ only by $\pm$ 0.3 mm/s around the mean of the four experiments: 20.9 mm/s. 

While the exact value of the fit parameter $v_0$ depends on the stochastic fluctuations within the granular gas during the heating phase, the value of $\tau$ can be compared to its prediction from granular kinetic theory \cite{vanNoije1998,Brilliantov2003}

\begin{equation}
\tau_{k.t.} = [\frac{\sqrt{2\pi}}{3}\chi(\phi)(1+\frac{3}{16}a_2)(1-\epsilon^2)nd^2\cdot v_{T0}]^{-1}.
\label{eq:tau_predict}
\end{equation}

Here $\chi(\phi)$ is the contact value of the pair correlation function. We use the usual approximation for hard sphere systems \cite{Carnahan1969}: $\chi(\phi)=(2-\phi)/(2(1-\phi)^3)$. $a_2$ is the second Sonine coefficient: $a_2=16(1-\epsilon)(1-2\epsilon^2)/(81-17\epsilon+30\epsilon^2(1-\epsilon))$. $v_{T0}$ is the thermal velocity of the particles at the beginning of the cooling: $v_{T0}=2/\sqrt{\pi}\cdot v_0$ (see appendix \ref{APP:2} for details). Using the fitted value $v_0$ = 20.6 mm/s from our first experiment, Eq. (\ref{eq:tau_predict}) predicts $\tau_{k.t.}$ = 1.34s, which is a factor of 2.6 times larger than the fitted value $\tau=$ 0.507s. 

The fact that $\tau<\tau_{k.t.}$
 matches with the limits of Eq. (\ref{eq:tau_predict}) 
which includes only the energy dissipation in the normal direction of the collisions, and not dissipation in tangential direction which can e.g.~be caused by inter-particle friction. Furthermore, inclusion of tangential dissipation means the rotational degrees of freedom join the dissipation mechanism and couple with the translational motion \cite{brilliantov:07}. 
We expect  that theories for rough particles \cite{luding:98,kranz:09,santos:11,bodrova:12} provide predictions closer to the experimental results. However, such a comparison, would also require 3D particle tracking including the rotational motion of the particles. Therefore this question depicts the direction of future experimental improvements.

\begin{figure}[ht]
\centering
\includegraphics[width=0.5\textwidth]{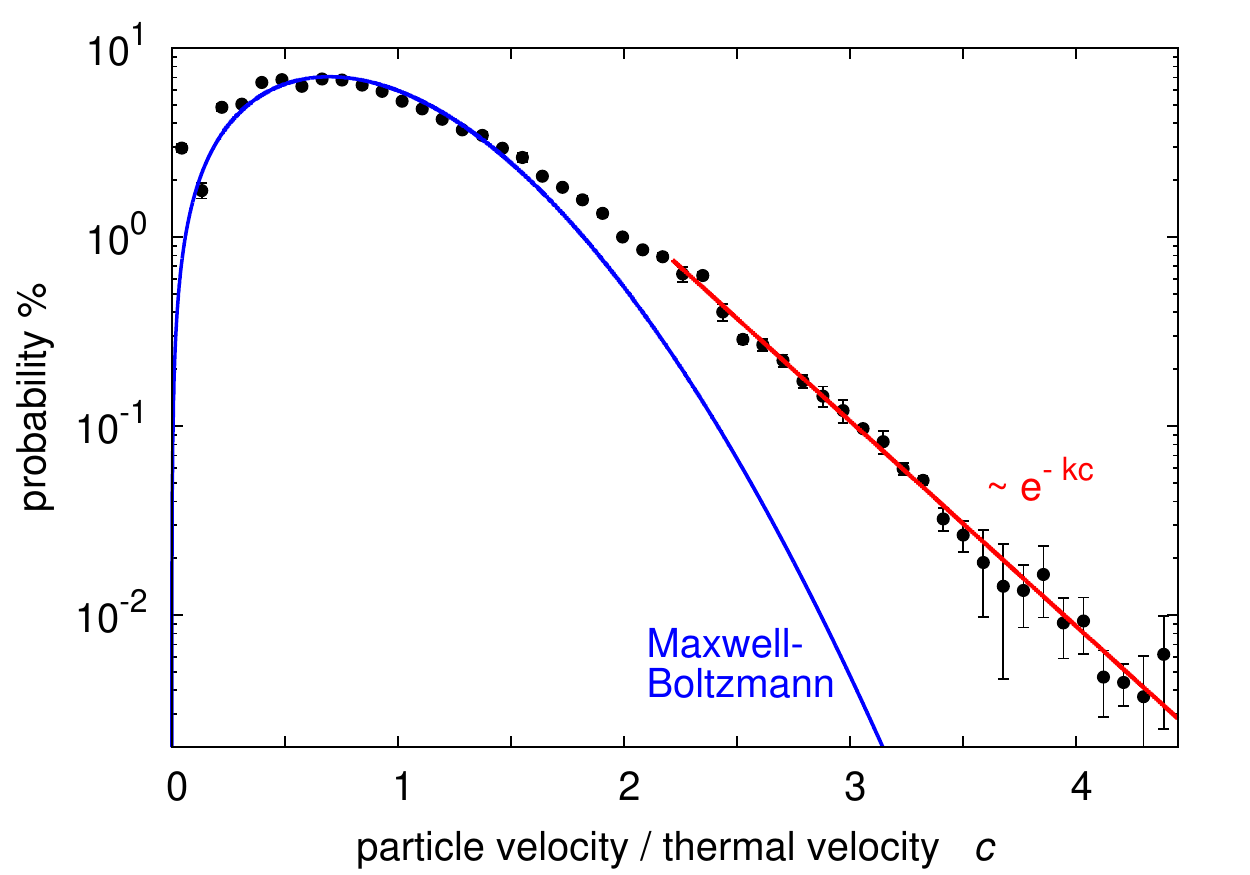}
\caption{The distribution of particle velocities in the homogeneous cooling state 
displays an exponential tail for $c>2.2$. 
Data is averaged over all four experiments shown in
Fig.~\ref{fig:4xHaff} and the time interval 1.8 to 3 seconds after onset of cooling. 
}
\label{fig:vel_dist_exponential}
\end{figure}

\textit{Velocity Distribution}.--
Fig.~\ref{fig:vel_dist_exponential} shows the distribution of the rescaled particle velocities $P(c)$. The high reproducibility of the experiments allows us to improve the statistics by averaging the distributions of the four experiments; the four individual distributions can be found in appendix \ref{APP:4}. For  values of $c$ smaller than 1.5, $P(c)$ can be approximated by a Maxwell-Boltzmann distribution. However, the main result of Fig.~\ref{fig:vel_dist_exponential} is that the high energy tail of $P(c)$ decays exponentially, as predicted by the kinetic theory of granular gases. A nonlinear fit 
of the experimental data using $P(c) \sim \exp(- k c^\alpha)$ results in $\alpha$ = 0.96 $\pm$ 0.19, which is again in good agreement with kinetic theory. 

Due to the intermittent nature of the magnetic driving (i.e. 20 ms on, 80 ms off), the velocity distribution of the heated gas, which is shown in appendix \ref{APP:5}, is different from a Maxwell-Boltzmann distribution and it does also not display a high velocity tail proportional to $\exp(- k v^{3/2} )$ as predicted by kinetic theory for the homogeneously driven gas. A fit of the tail with $P(c) \sim \exp(- k c^\alpha)$ results in $\alpha$ = 0.72 $\pm$ 0.21.

The fact that the system still relaxes towards the distribution predicted by granular kinetic theory can be interpreted as the equivalent of Boltzmann's H-theorem for thermal systems \cite{Chapman1960}. It is essential for the comparison with theoretical predictions that the observed velocity histogram allows for the interpretation as a distribution function including the definition of a well-defined temperature. 

\textit{Conclusions.--}We have shown experimentally that the  kinetic theory of granular gases predicts correctly the two most fundamental properties of a freely-cooling three-dimensional gas: the evolution of the average particle velocity and the appearance of an exponential tail in the particle velocity distribution. Building on this foundation, we expect quantitative deviations between experiment and existing theory to stimulate further fruitful work.

We acknowledge partial funding from BMWi/DLR through project 50WM1651 and 50WM1945. 
We also thank an anonymous referee for providing valuable insights into the theoretical implications of our data.

\clearpage

\begin{appendices}

\appendixtitleon
\renewcommand{\thesubsection}{\Alph{section}.\arabic{subsection}}

\section{Particle tracking}
\label{APP:1}
\begin{figure}[ht]
\centering
\includegraphics[width=0.95\linewidth]{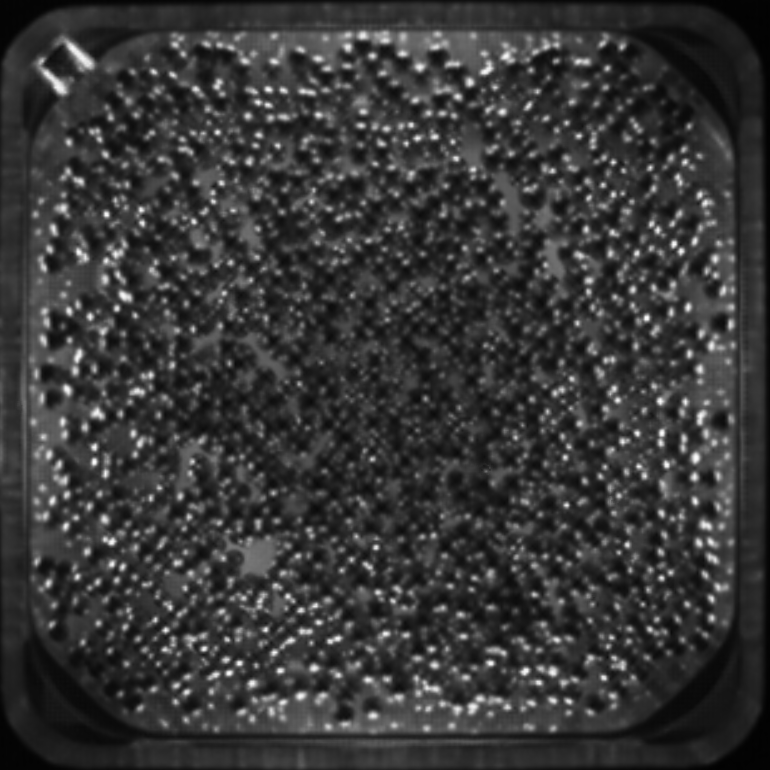}
    \caption{The starting point is a total focus image with 770 x 770 pixels. 
    }
    \label{fig:raw}
\end{figure}

Fig.~\ref{fig:raw} displays a typical raw images from which we track the motion of the visible
subset of spheres. Usually, Particle Tracking Velocimetry (PTV) identifies in a first step 
in each image the positions of all visible particles. In a second step the lists 
of particle coordinates originating from subsequent images are used to construct self-consistent particle trajectories \citep{ouellette:06,jaqaman:08,besseling:09}.

This strategy will not work for our raw data because the identification of the particles in an individual image is hampered by three experimental constraints: a) the intensity contrast between particles and background is small. b) the four illumination sources result in a spatially dependent number of reflections on each sphere which prohibits the use of template matching algorithms. And c) the relatively high particle density results in frequent partial occlusion of spheres.

\begin{figure}[h]
\centering
\includegraphics[width=0.95\linewidth]{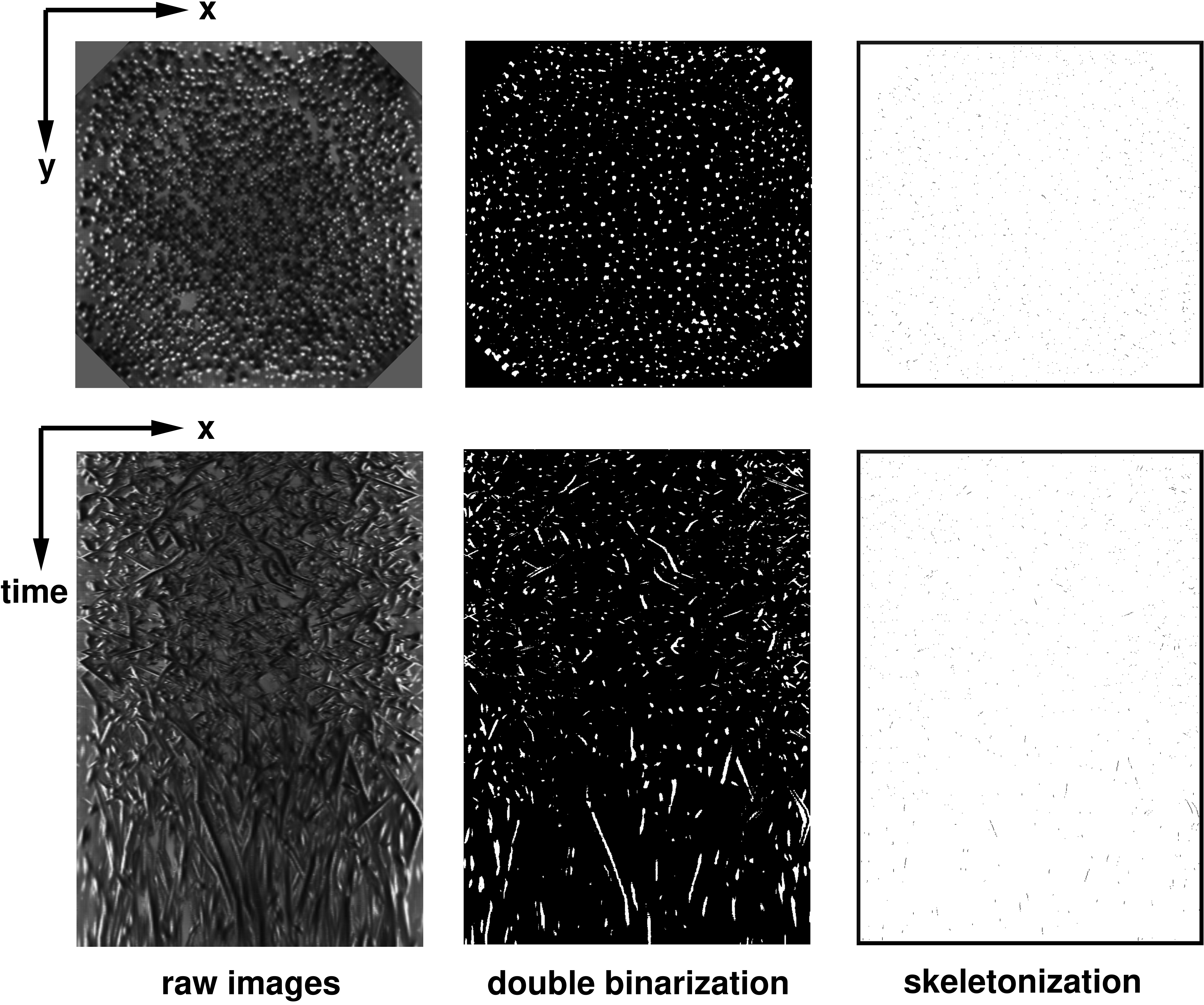}
    \caption{Preprocessing of an image sequence. The \textit{top row} contains individual images, 
    the \textit{bottom row} displays a cut through the center of the three dimensional image stack;
    this view corresponds to a space-time plot.
    The \textit{left column} shows the raw images. The four corners of the image are cut out 
    as they contain only the bevels in the corners of the experimental cell.
    The space-time view reveals the particle motion (in x direction) as inclined bright and dark
    streaks (originating from the reflections and shadows on the sphere surfaces). The smaller the 
    angle between a streak and the vertical axis is, the slower the corresponding particle moves.
    The \textit{middle column} shows the result of a double binarisation with an upper
    and a lower threshold: the streaks are now separated from the background.
    The \textit{right column} corresponds to the middle column after the streaks have been reduced to lines of width one pixel using a 3D skeletonization. Images have been inverted for better 
    visibility.
    }
    \label{fig:preprocessing}
\end{figure}

In order to avoid these problems, we have decided to merge identification and tracking into a single step, i.e., to use multiple consecutive images to identify where our particles are. The basic idea can be seen in the left column of Fig.~\ref{fig:preprocessing}. The image in the top row is a cropped version of the image in Fig.~\ref{fig:raw}. The image in the bottom row is a space-time plot which has been constructed as follows: take all 1000 images corresponding to a given experimental run (heating phase and then free cooling) and stack them on top of each other, effectively creating a 3D volume image. Then we cut a single vertical slice from this volume. Which shows how a specific row in one of our images evolves with time (i.e.~going from top to bottom in that slice.)

The bright streaks in this space-time image correspond to reflections on the sides of the spheres, the dark streaks are the dark areas at the centers of the spheres (because there is no light source at the camera side of the cell.) Streaks start and end because particles also have velocity components in the y-direction, i.e.~perpendicular to the single line we are visualizing here. In some cases the ends of the streaks will also correspond to particles appearing or becoming hidden behind other particles.

But the main point here is that the extended nature of this streaks allows us to identify moving entities together with their trajectory. While the streaks themselves will not necessarily correspond to the particle centers, the symmetric and diffuse illumination guarantees 
that we do not make a parallax error if we identify the streak velocity with the particle 
velocity. In consequence, this method is not suitable to measure pair correlation functions or
mean square displacements, but it will produce reliable velocity distributions and mean velocities. 

The actual image processing pipeline consist of two steps: \textit {preprocessing} which creates 
pixel-thin representations of the streaks and \textit{track identification} which detects 
the streaks and converts them into particle velocities. These steps are further described in the next two subsections.

\subsection{Preprocessing}
Before assembling all images of a given experiment into a 3D image stack, as shown in the bottom left panel of Fig.~\ref{fig:preprocessing}, we crop them to remove all pixels belonging to the cell boundaries. 

The next step is to turn the spatio-temporal paths of the reflections and central shades (on the spheres) into separate objects. This is achieved with a double binarisation where we choose the upper and lower threshold such that the darkest 2.9 \% and brightest 1.8 \% of pixels are converted to white while the remaining background is set to black. The middle column of Fig.~\ref{fig:preprocessing} provides an example. Changing these thresholds such that the number of pixels retained varies by a factor of 1.5 changes the $\tau$ values of the Haff's law fits by less than $\pm$6\%. Which is well within the variance between different experimental runs.

The final step is a reduction of the white objects to lines of thickness one pixel. This is achieved using a skeletonization algorithm with a 27 pixel neighborhood in 3D (Skeletonization works by making successive passes through the volume. On each pass, border pixels, i.e.~white pixels with at least one black neighbor, are identified and removed on the condition that this does not break the connectivity of the skeleton). The right column of Fig.~\ref{fig:preprocessing} displays the skeleton of our example image: the streaks representing our particles have now a diameter of one pixel. (The lines appear shorter because  most particles have also a velocity component in y-direction which makes the skeleton lines move out of the the specific slice shown here.)

\subsection{Identifying tracks}
In micro-gravity particles will move between collisions with constant speed. In our spatio-temporal image stack constant speed translates into straight lines; we therefore want to fit our streaks with straight line segments.

Identifying straight lines in an image is normally done with the so-called Hough transformation which introduces an additional accumulator space with axis which are  parameters describing all possible lines. For 2D images these parameters are normally the pair of angle $\theta$ and distance $r$ which constitute the Hesse Normal form of a straight line. Then for each white pixel the $\theta$ and $r$ pairs of all lines possibly going through this pixel are computed and the corresponding bins in the accumulator space are incremented by one. Once all white pixels have been processed, the maxima in the accumulator space indicate the parameters of the lines on which many white pixels can be found.

\begin{figure}[ht]
\centering
\includegraphics[width=0.95\linewidth]{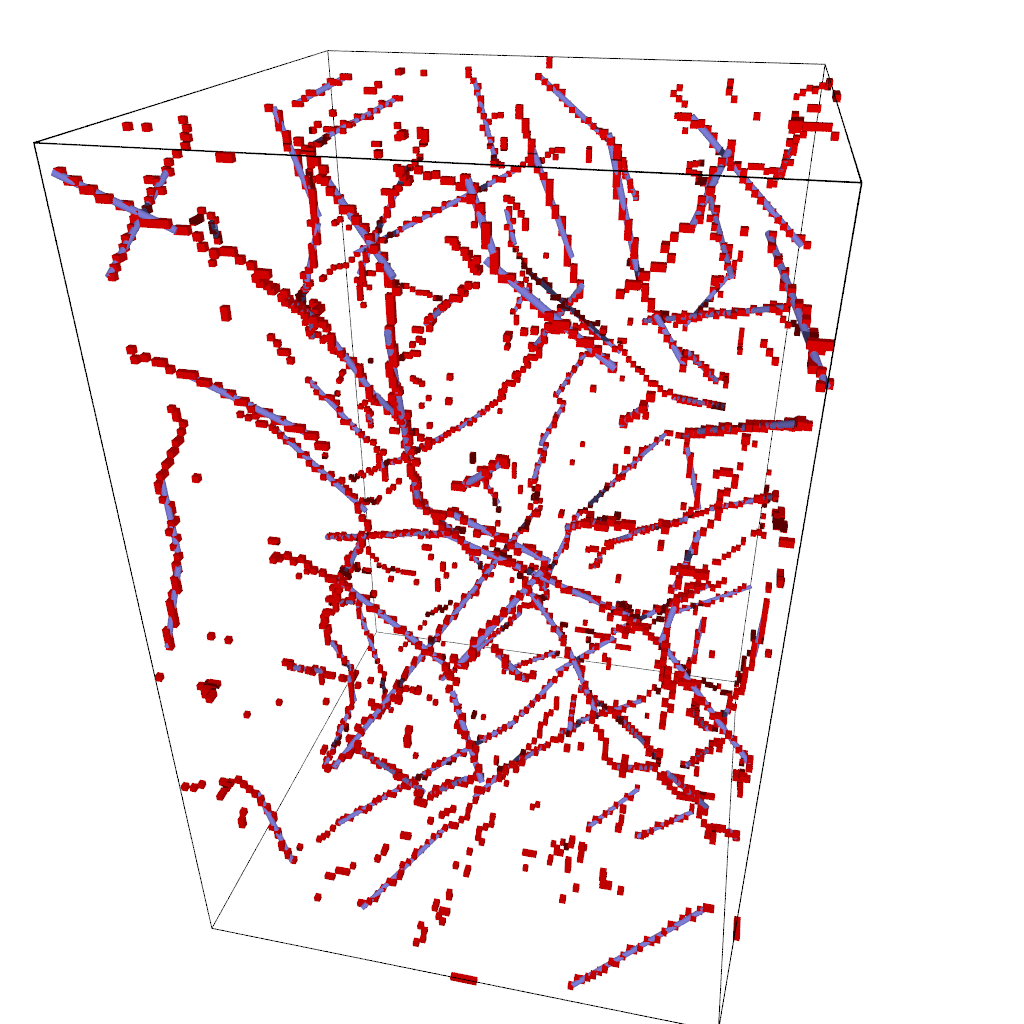}
    \caption{Identifying straight line segments using 3D Hough transformation.
    Data are a subset of the data shown in Fig.~\ref{fig:preprocessing}, right column.
    The volume inside the black wire-frame corresponds to a volume of 100 by 100 voxels
    (horizontal dimensions), tracked over 150 time steps/images (vertical dimension).
    Red cubes are the pixels identified by the skeletonization as the backbone of our streaks.
    Purple cylinders represent the result of the 3D Hough transformation. An animated version of 
    this figure can be found in the supplemental material.  
    }
    \label{fig:haystack}
\end{figure}

Generalizing this idea to 3D requires a number of modifications and expansions. Luckily, most of this work has already been done by Dalitz and coworkers (2017). Their paper is also accompanied by a well documented, open-source implementation in C named \texttt{houdh3dlines}. \nocite{dalitz:17} Fig.~\ref{fig:haystack} demonstrates how we can use this 3D Hough transformation to identify the line segments which form the backbone in the center of the streaks.

The parameterization in 3D requires 4 parameters: two angles to describe the direction of the line and two components of a vector going from the origin of the coordinate system to the intersection point between the line and a plane which is perpendicular to that line (and also going through the origin). For a typical experimental run with an image stack of size 700 times 700 pixels times 1000 time steps this translates into a memory requirement of 10.1 GByte for the accumulator space. 

After inserting the information contained in the white pixels of our 3D volume into this accumulator space, \texttt{houdh3dlines} returns the parameters of all maxima in that space in the sequence of decreasing bin height, i.e., decreasing line length. It also performs a conventional line fit to the points constituting each maximum in order to increase the accuracy of the four parameters describing the line.

We have slightly modified \texttt{houdh3dlines} for our purpose: we only retain line segments which are based on streaks with white pixels in at least six consecutive time steps. Moreover, we also record the earliest and latest frame contributing to that line segment. From the angle between the line direction vector and the time axis we can compute the x and y component of the velocity of the underlying particle. In the subsequent analysis this velocity is considered to be present between the start and end time of this line segment.

\section{Relationship between measured average velocity $\langle v \rangle$ and the thermal velocity $v_{T}$} 
\label{APP:2}

It is clear from the previous section that the particle velocity we measure is the magnitude of the velocity vector projected from a three-dimensional volume into a two-dimensional image plane.
Therefore, the relevant Maxwell-Boltzmann (MB) distribution becomes  the 2D version: 
$P(v) \sim v \cdot \exp[-v^2/v_T^2]$. 
After integration we find 
$v_{T}= 2/\sqrt{\pi}\cdot \langle v \rangle$ (If we would use the 3D version of the MB distribution, $P(v) \sim v^2\cdot \exp[-v^2/v_T^2]$, we would obtain the opposite relationship:$\langle v \rangle = 2/\sqrt{\pi}\cdot v_{T}$).

For the calculation of the Haff's time (Eq.2 of the paper) we need the thermal velocity 
$ v_{T0} = v_T|_{t=0} $
at the begin of the cooling.  
As described above, $ v_{T0}$ can be computed from 
the mean velocity measured at the starting of the cooling $v_0=\langle v\rangle |_{t=0}$  by $v_{T0}=2/\sqrt{\pi}\cdot v_0$. 

\section{Clustering after 4 seconds}
\label{APP:3}

\begin{figure}
\begin{subfigure}{.49\linewidth}
\centering
  \includegraphics[width=\linewidth]{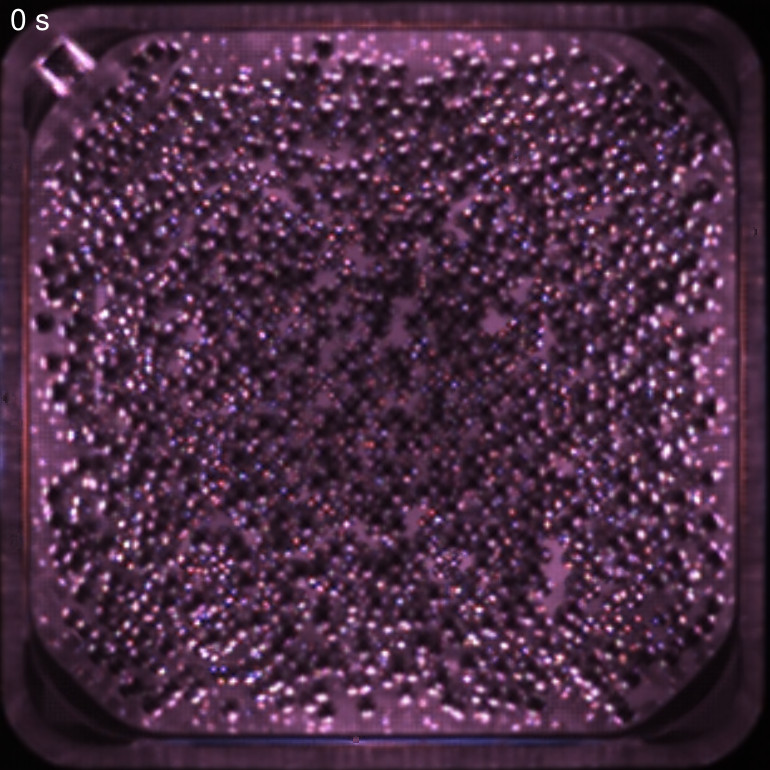}
\end{subfigure}
\begin{subfigure}{.49\linewidth}
  \centering
  \includegraphics[width=\linewidth]{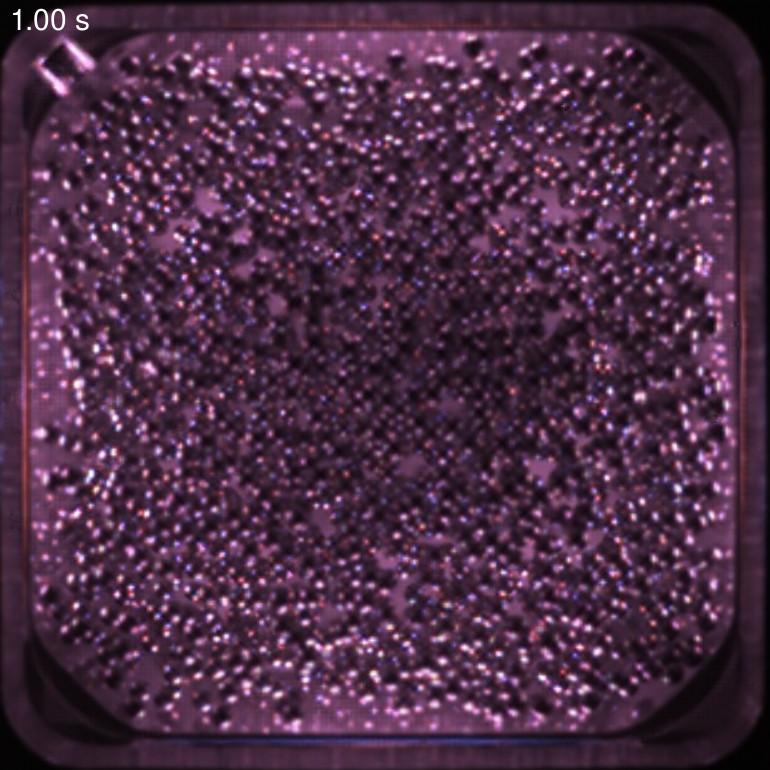}
\end{subfigure}

\begin{subfigure}{.49\linewidth}
\centering
  \includegraphics[width=\linewidth]{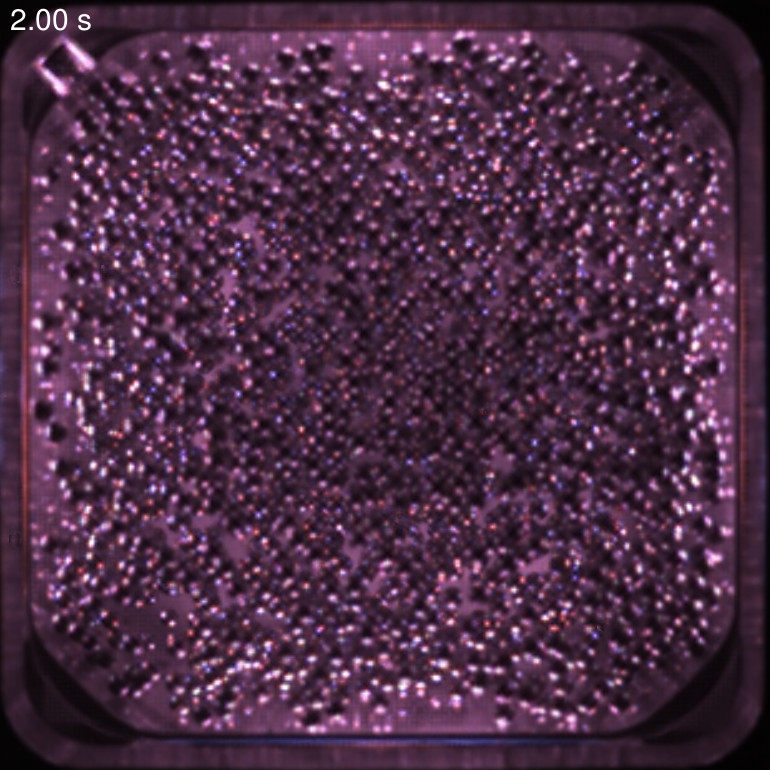}
\end{subfigure}
\begin{subfigure}{.49\linewidth}
  \centering
  \includegraphics[width=\linewidth]{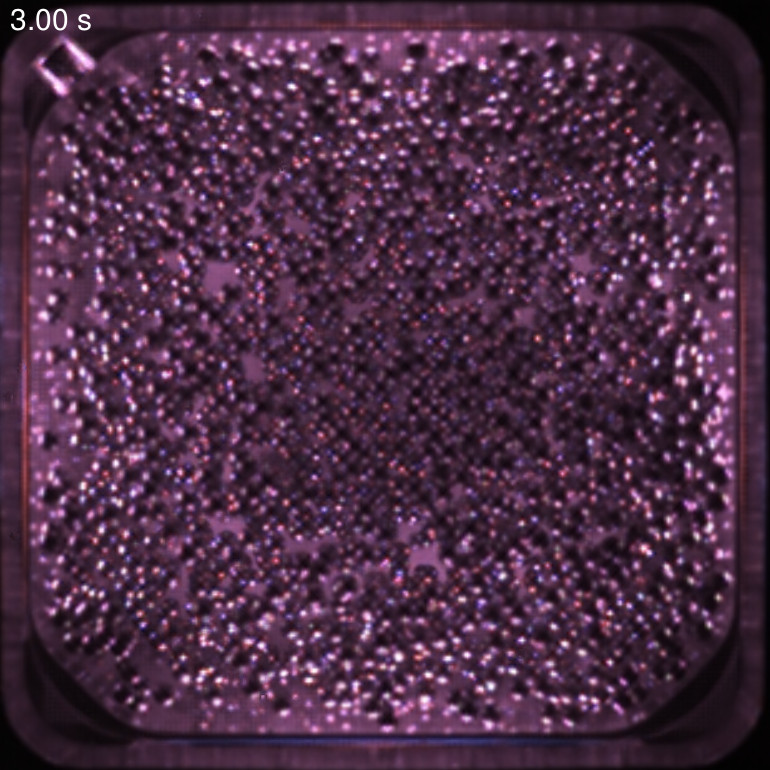}
\end{subfigure}

\begin{subfigure}{.49\linewidth}
\centering
  \includegraphics[width=\linewidth]{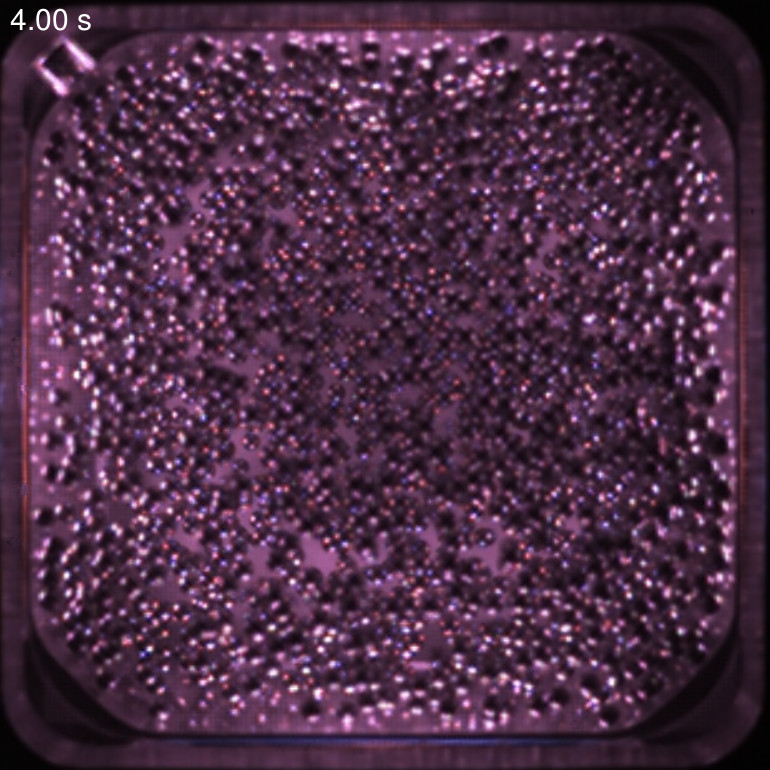}
\end{subfigure}
\begin{subfigure}{.49\linewidth}
  \centering
  \includegraphics[width=\linewidth]{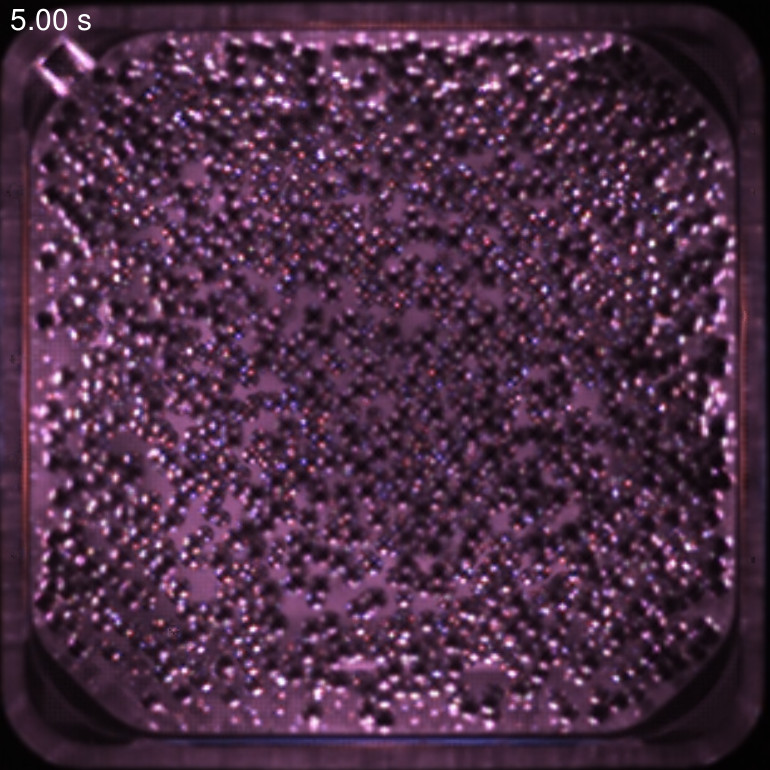}
\end{subfigure}
\caption{The first 5 seconds of the cooling phase of the first experiment.}
\label{fig:first_cooling_1}
\end{figure}

\begin{figure}

\begin{subfigure}{.49\linewidth}
\centering
  \includegraphics[width=\linewidth]{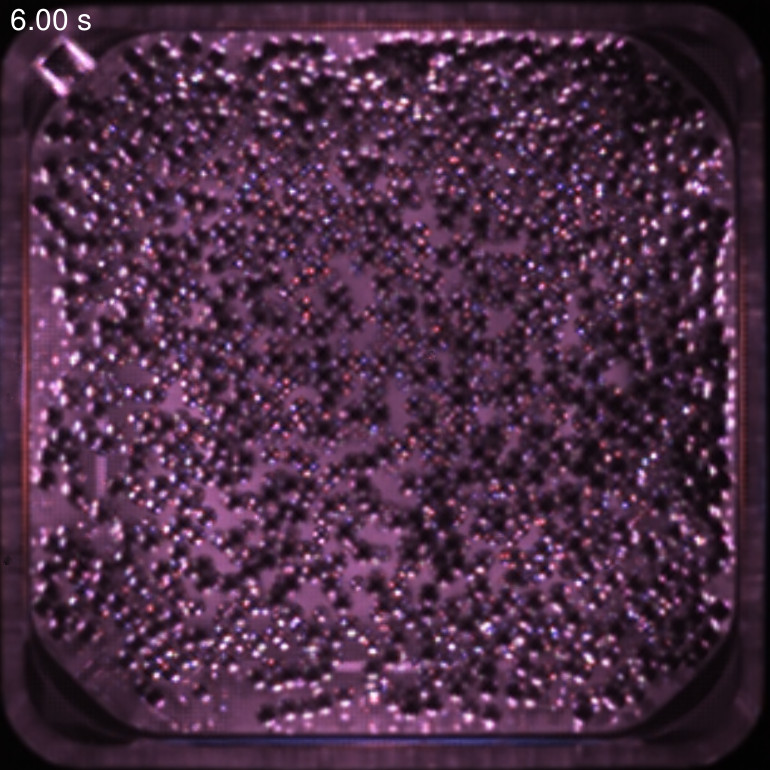}
\end{subfigure}
\begin{subfigure}{.49\linewidth}
  \centering
  \includegraphics[width=\linewidth]{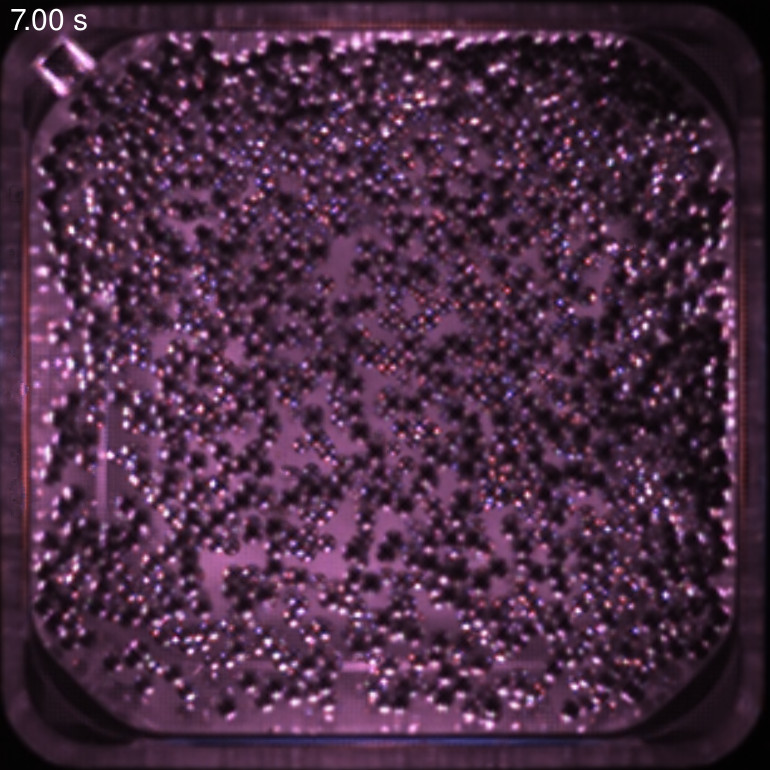}
\end{subfigure}

\begin{subfigure}{.49\linewidth}
\centering
  \includegraphics[width=\linewidth]{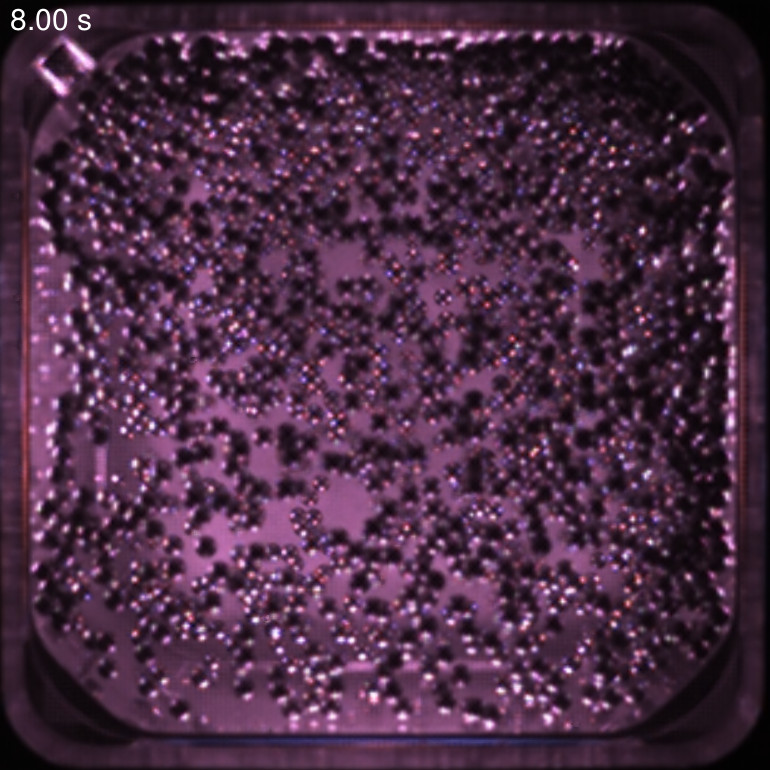}
\end{subfigure}
\begin{subfigure}{.49\linewidth}
  \centering
  \includegraphics[width=\linewidth]{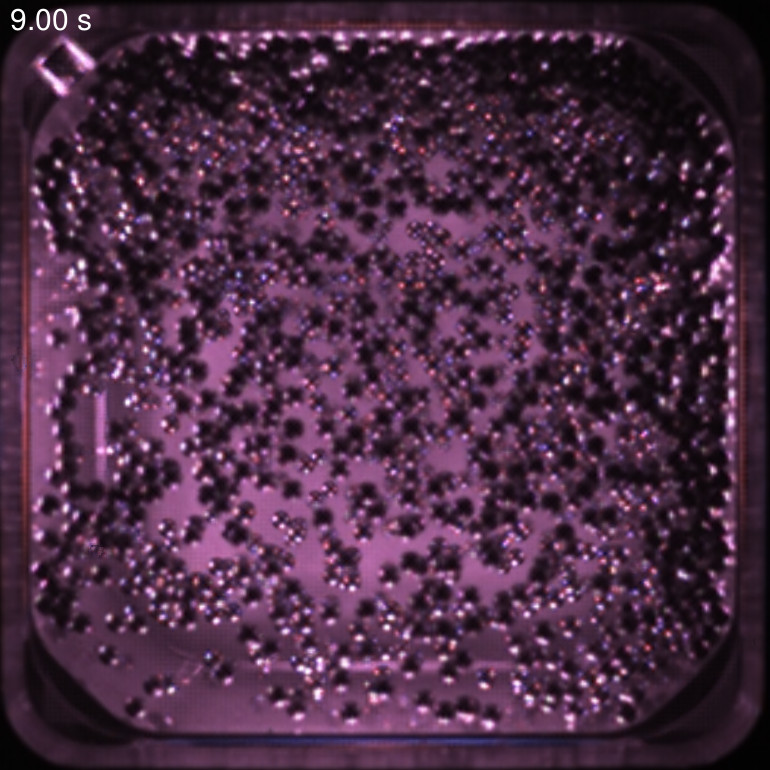}
\end{subfigure}
\caption{Continuing figure \ref{fig:first_cooling_2}, the next four seconds of the cooling phase of the first experiment.}
\label{fig:first_cooling_2}
\end{figure}

The spatial homogeneity of the particle distribution is a consequence of the magnetic driving. Once the gas enters the free cooling phase, the remnant gravitational acceleration will start to create inhomogeneities. Figures \ref{fig:first_cooling_1} and \ref{fig:first_cooling_2} demonstrate that denser cluster begin to form at the "top" and "right" edge of the sample cell after approximately 4 seconds.  

Based on figures \ref{fig:first_cooling_1} and \ref{fig:first_cooling_2}, we can try to estimate the remnant acceleration. If we assume that particles migrate about 1 cm in 8 s, the remnant acceleration would correspond to $3 \times 10^{-5} g$, which is in agreement with the specification of the sounding rocket flight.

\section{Velocity distributions of individual runs} 
\label{APP:4}
\begin{figure}[ht]
\centering
\includegraphics[width=0.95\linewidth]{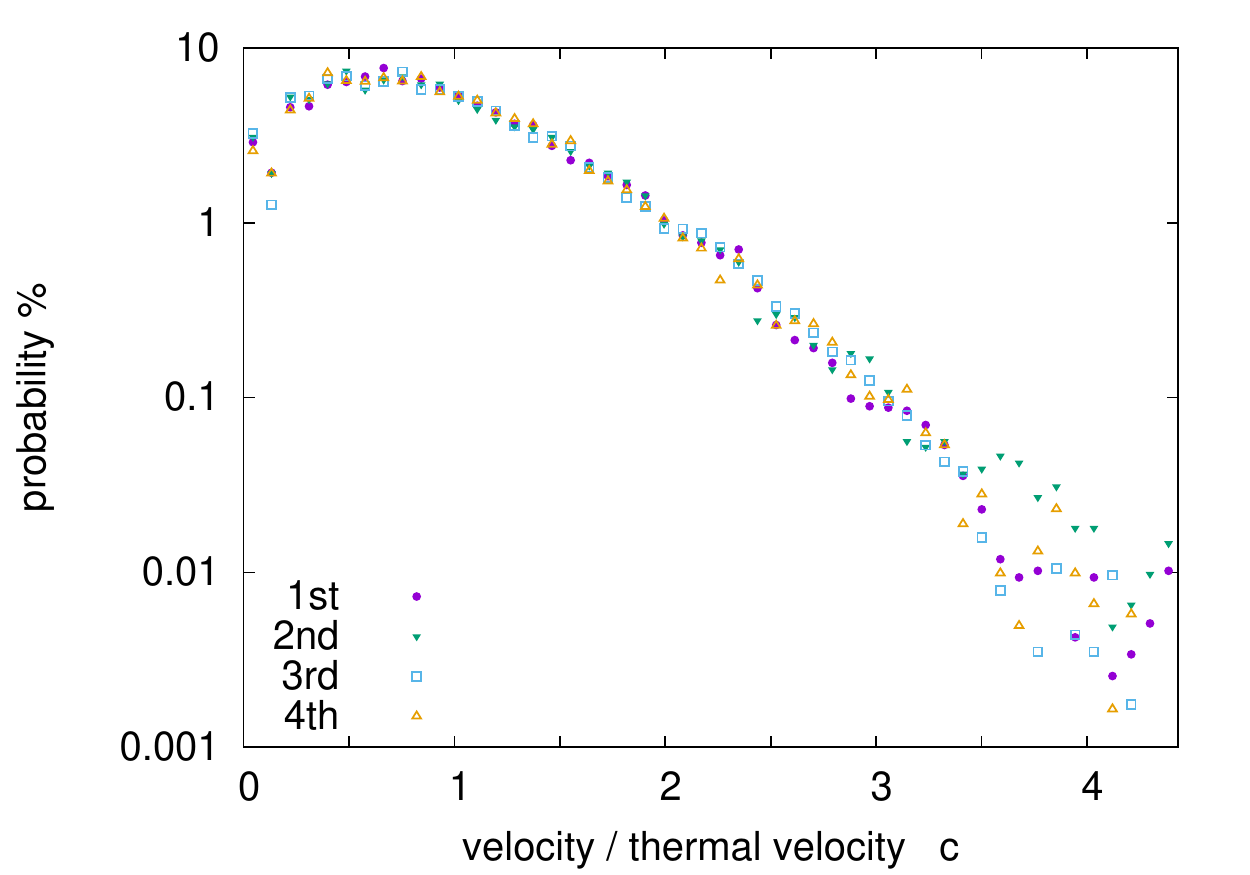}
\caption{The rescaled velocity distributions of all four experiments,
measured 1.8 to 3 seconds after onset of cooling. 
}
\label{fig:4xvelocity_distributions}
\end{figure}

Fig.~\ref{fig:4xvelocity_distributions} shows the rescaled velocity distributions of the four 
individual experiments. The average of the four distribution is shown in Fig.~3 of the paper.

\section{Properties of the heated granular gas}
\label{APP:5}
It has been pointed out, based on numerical simulations, that the evolution of the velocity distribution from a normal Maxwell-Boltzmann distribution towards the solution predicted by kinetic theory can take up to the order of ten times $\tau$ \citep{poeschel:07}.

\begin{figure}[ht]
\centering
\includegraphics[width=0.9\linewidth]{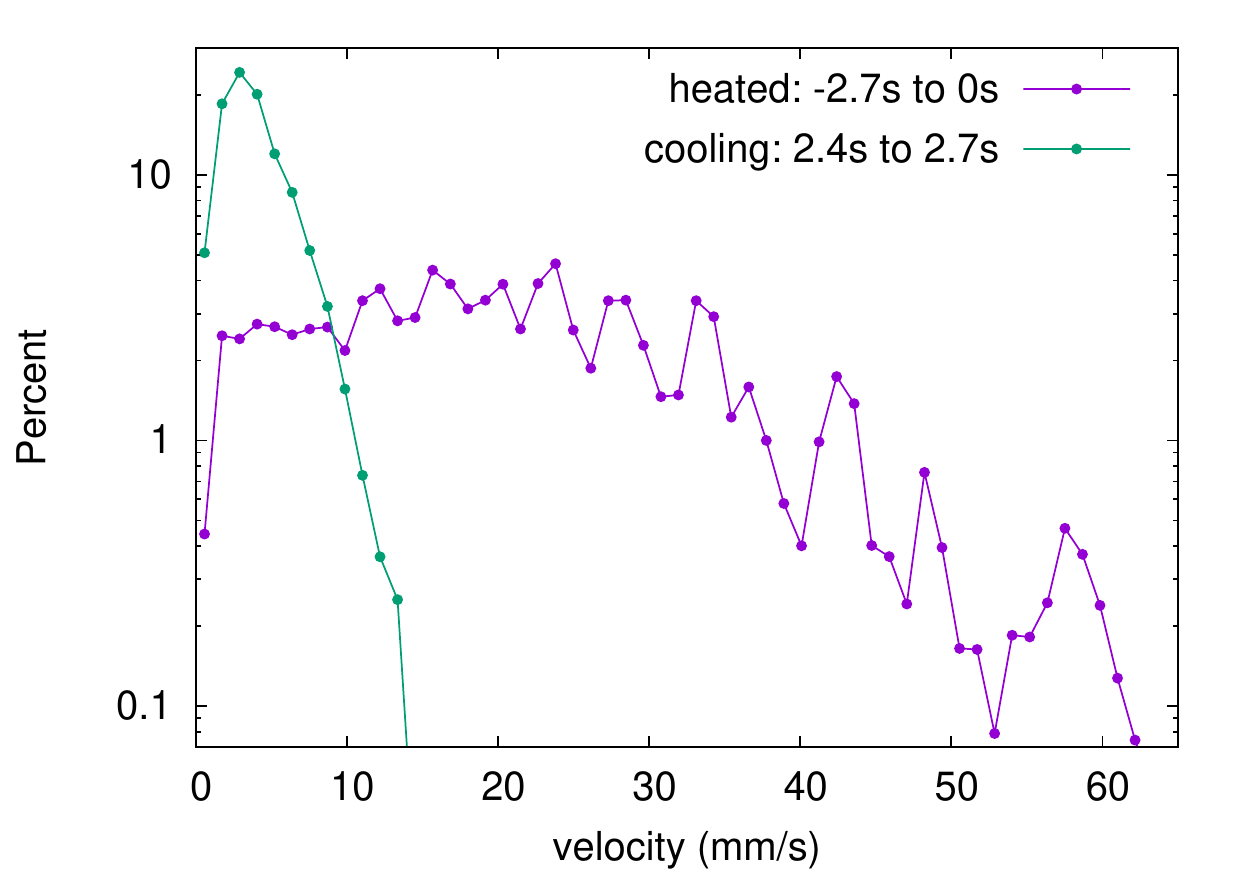}
\includegraphics[width=0.9\linewidth]{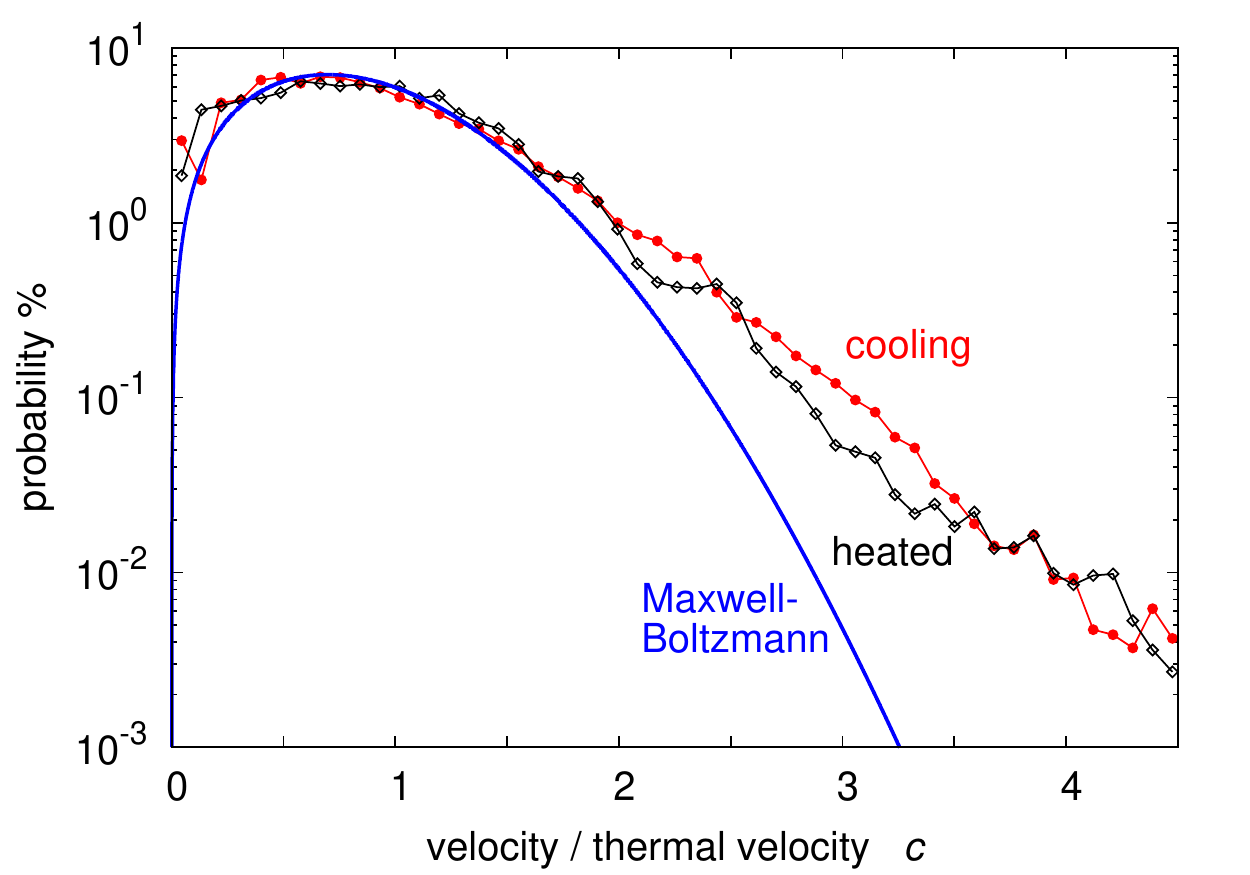}
\caption{Velocity distributions of the heated and the cooling state of the granular gas.
\textit{Upper panel:}
The non-rescaled velocity distributions in the heated and cooled case differ significantly. Data is from experiment 1, the cooling phase is analyzed for only 0.3 s in order to minimize distortions by the ongoing decrease of the thermal velocity.
\textit{Lower panel:}
Rescaled velocity distribution, 
averaged over all four experiments. The heated case has a higher probability of particles with velocities in the range $1<c<1.5$ and also a different exponent of the tail, as shown in figure 
\ref{fig:exponent_fit}. However, the heated velocity distribution is clearly more similar to the cooling distribution than a Maxwell-Boltzmann
distribution.
}
\label{fig:_vel_dist_comparison}
\end{figure}

We find in our experiment that the fully developed velocity distribution can already be found after three times $\tau$. This is a beneficial side effect of the 1:4 duty cycle of our driving protocol; during the 80 ms where the magnets are switched off, the system can already start to relax towards the HCS. This becomes evident from Fig.~\ref{fig:_vel_dist_comparison} where the velocity distribution in the heated gas is already approaching the HCS when compared to the Maxwell-Boltzmann distribution.

\begin{figure}[hb]
\centering
\includegraphics[width=0.9\linewidth]{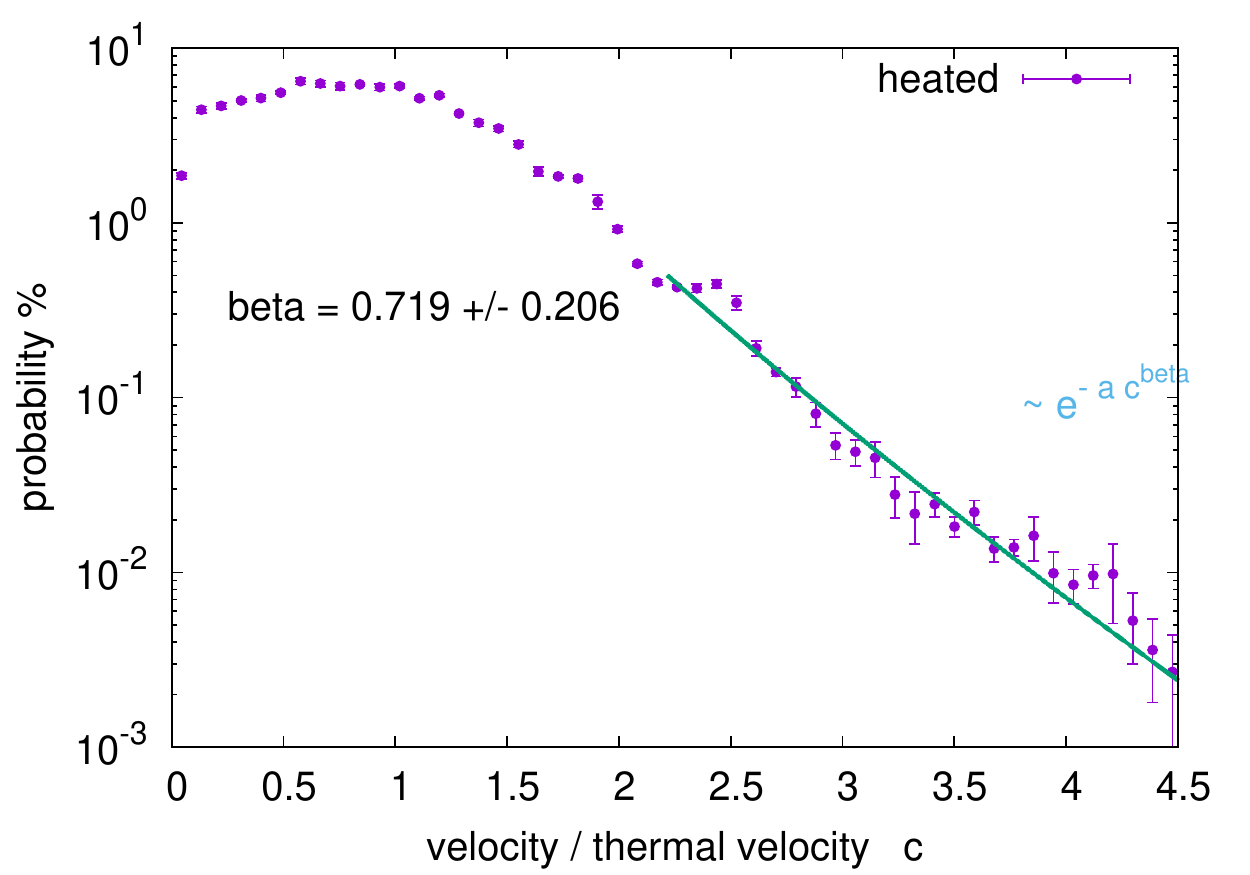}
\includegraphics[width=0.9\linewidth]{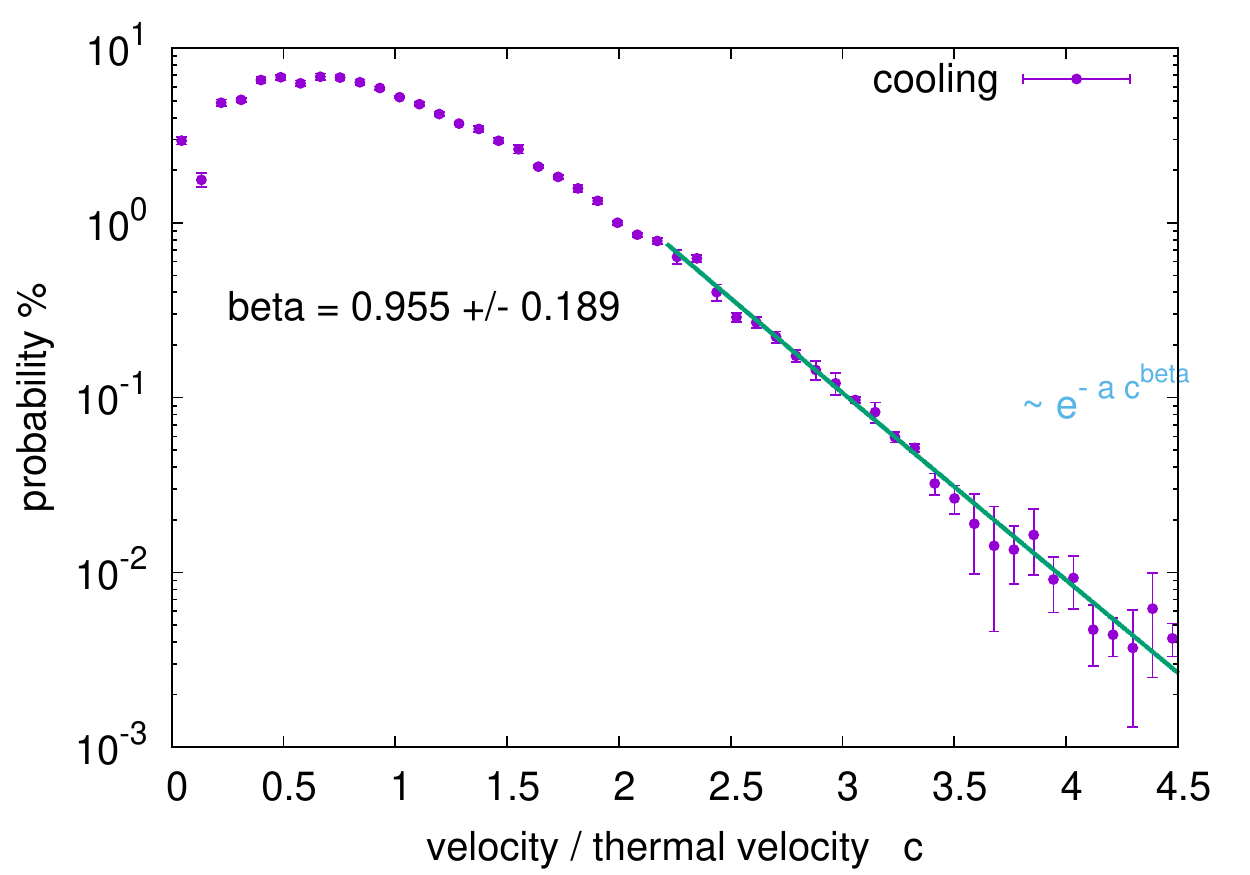}
\caption{Fit of $P(c) \sim \exp(- k c^\beta)$ to the velocity distributions of both the heated (\textit{Upper panel}) and cooling (\textit{Lower panel})
phase of the experiment. 
Data is averaged over all four experiments.
The time interval for the cooling case is 1.8s to 3s
after switching of the magnets, for the heated case it is the 2.2 (2.7 for experiment 1) seconds before. }
\label{fig:exponent_fit}
\end{figure}

However, there are still significant differences between the heated and cooling velocity distribution; especially in the shape of the high velocity tail:
Figure \ref{fig:exponent_fit} shows that the tail of the heated distribution can not be described by an  exponential decay.\\

\section{The cooling  gas reaches a stationary state}
\label{APP:6}
\begin{figure}[tbp]
\centering
\includegraphics[width=0.8\linewidth]{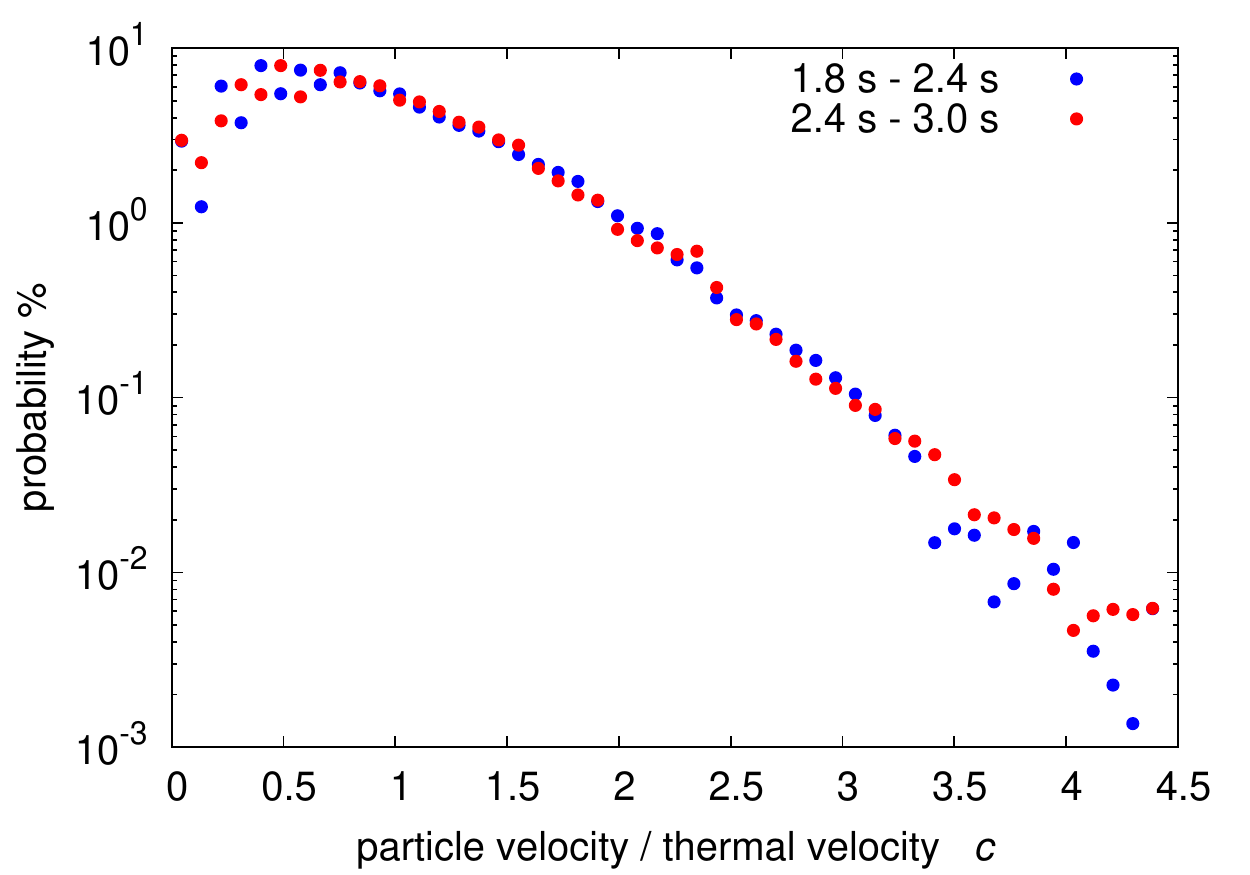}
\caption{The re-scaled velocity distributions during cooling, averaged over all four experiments, but split into two time intervals. Within statistical fluctuations the two distributions are identical, proving the point that the distribution is stationary.}
\label{fig:early_late}
\end{figure}

Given the limited time frame of cooling before clustering sets in, the question could be raised if we measure indeed the asymptotic distribution. 
This question is answered in figure \ref{fig:early_late} where we have split the time range shown in figure 3 in the paper (1.8 to 3 s) into two halves. Figure \ref{fig:early_late} then demonstrates that the  velocity distribution does not change between these two intervals which corroborates the existence of a steady state.\\

\section{The cooling time scale for viscoelastic particles}
\label{APP:7}

\begin{figure}[tbp]
\centering
\includegraphics[width=1\linewidth]{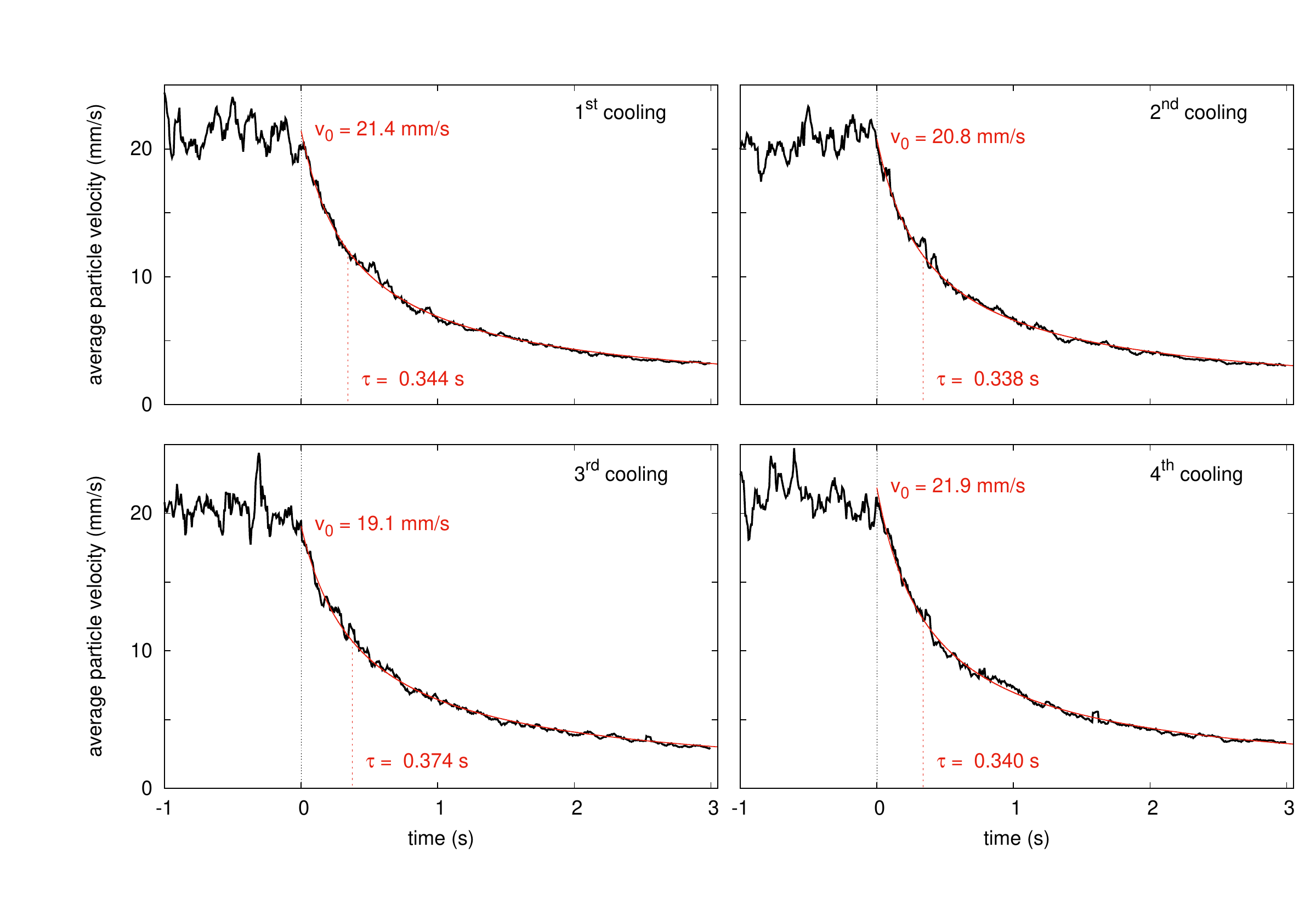}
\caption{Fitting the viscoelastic velocity decay to all four experiments}
\label{fig:4xviscoelastic}
\end{figure}

As mentioned in the manuscript, in addition to the Haff's law for constant $\epsilon$, that for $\epsilon=\epsilon(g)$, i.e.,  $\langle v(t)\rangle = v_0/ (1+t/\tau)^{5/6}$ for viscoelastic particles, has also been used to fit the experimental cooling curves (see Fig. \ref{fig:4xviscoelastic}). The resulting time scale $\tau$ can then be compared with theories.\\

Eq. 3.22, 9.14 and 9.29 of \citep{Brilliantov2003} establishes how $\tau$ depends on the viscoelastic collision property $\epsilon=\epsilon(g)$. We rewrite them using the conventions in the main manuscript.
\begin{equation}
\label{eq:1}
\begin{aligned}
&\epsilon=1-C_1A\kappa^{2/5}g^{1/5}+C_2A^2\kappa^{4/5}g^{2/5}+...,\\
&\delta=A\kappa^{2/5}(\frac{1}{2}v_{T0}^2)^{1/10},\\
&\tau_{v.s.}^{-1}=\frac{64q_0}{5}\sqrt{\pi}\chi(\phi)d^2n\sqrt{\frac{v_{T0}^2}{2}}\delta\\
\end{aligned}
\end{equation}
, where $q_0\approx0.173$, $C_1\approx 1.15334$ and $C_2\approx 0.79826$ are pure numbers, $v_{T0}$, $d$, $n$, and $\chi(\phi)$ remain the same as used in the main manuscript. The only unknown number $A\kappa^{2/5}$ can be found from our lab calibration experiment mentioned in the main manuscript.\\

\begin{figure}[tbp]
\centering
\includegraphics[width=0.95\linewidth]{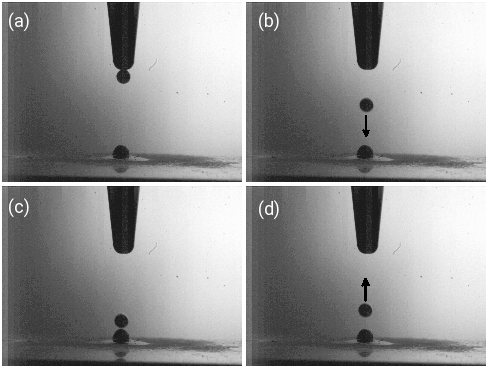}
\caption{Snapshots of the lab measurement of the coefficient of restitution}
\label{fig:gexp_CoR}
\end{figure}

As illustrated in Fig. \ref{fig:gexp_CoR}, $\epsilon$ used in the main manuscript is retrieved from a simple particle dropping experiment. Two MuMetall particles from the same batch used in the 0g experiment are prepared, with one completely fixed by glue to a solid surface below, and another temporarily attached to a nozzle above (a). The nozzle's position is finely adjusted, so that when the particle above drops (b), it collides with the one below almost vertically (c). This close-to-vertical collision minimizes the rotation after the particle above is bounced back (d). The translational velocities before and after the collision are calculated from image analysis and lead to the result of $\epsilon=0.66$ for a collision velocity $g=368mm/s$.\\

For viscoelastic particles, $\epsilon(g=368mm/s)=0.66$ fully determines $A\kappa^{2/5}$ in the first line of Eq. \ref{eq:1}. The second and the third lines then predict, for our low-gravity cooling, a time scale of $\tau_{v.s.}=0.982s$.\\

Comparing with the viscoelastic fit result shown in Fig.~\ref{fig:4xviscoelastic}, we find that $\tau_{v.s.}$ is again more than 2.5 times larger than the experimental value (e.g., $\tau=0.344s$ from the 1st cooling). Therefore, viscoelastic collision theory does not interpret our experiments quantitatively better compared to the simpler kinetic theory assuming $\epsilon$ to be constant.
\end{appendices}

\bibliography{megrama}  

\begin{thebibliography}{83}%
\makeatletter
\providecommand \@ifxundefined [1]{%
 \@ifx{#1\undefined}
}%
\providecommand \@ifnum [1]{%
 \ifnum #1\expandafter \@firstoftwo
 \else \expandafter \@secondoftwo
 \fi
}%
\providecommand \@ifx [1]{%
 \ifx #1\expandafter \@firstoftwo
 \else \expandafter \@secondoftwo
 \fi
}%
\providecommand \natexlab [1]{#1}%
\providecommand \enquote  [1]{``#1''}%
\providecommand \bibnamefont  [1]{#1}%
\providecommand \bibfnamefont [1]{#1}%
\providecommand \citenamefont [1]{#1}%
\providecommand \href@noop [0]{\@secondoftwo}%
\providecommand \href [0]{\begingroup \@sanitize@url \@href}%
\providecommand \@href[1]{\@@startlink{#1}\@@href}%
\providecommand \@@href[1]{\endgroup#1\@@endlink}%
\providecommand \@sanitize@url [0]{\catcode `\\12\catcode `\$12\catcode
  `\&12\catcode `\#12\catcode `\^12\catcode `\_12\catcode `\%12\relax}%
\providecommand \@@startlink[1]{}%
\providecommand \@@endlink[0]{}%
\providecommand \url  [0]{\begingroup\@sanitize@url \@url }%
\providecommand \@url [1]{\endgroup\@href {#1}{\urlprefix }}%
\providecommand \urlprefix  [0]{URL }%
\providecommand \Eprint [0]{\href }%
\providecommand \doibase [0]{http://dx.doi.org/}%
\providecommand \selectlanguage [0]{\@gobble}%
\providecommand \bibinfo  [0]{\@secondoftwo}%
\providecommand \bibfield  [0]{\@secondoftwo}%
\providecommand \translation [1]{[#1]}%
\providecommand \BibitemOpen [0]{}%
\providecommand \bibitemStop [0]{}%
\providecommand \bibitemNoStop [0]{.\EOS\space}%
\providecommand \EOS [0]{\spacefactor3000\relax}%
\providecommand \BibitemShut  [1]{\csname bibitem#1\endcsname}%
\let\auto@bib@innerbib\@empty
\bibitem [{\citenamefont {Brilliantov}\ and\ \citenamefont
  {P\"oschel}(2004)}]{Brilliantov2003}%
  \BibitemOpen
  \bibfield  {author} {\bibinfo {author} {\bibfnamefont {N.}~\bibnamefont
  {Brilliantov}}\ and\ \bibinfo {author} {\bibfnamefont {T.}~\bibnamefont
  {P\"oschel}},\ }\href@noop {} {\emph {\bibinfo {title} {Kinetic Theory of
  Granular Gases}}}\ (\bibinfo  {publisher} {Oxford University Press},\
  \bibinfo {year} {2004})\BibitemShut {NoStop}%
\bibitem [{\citenamefont {Kadanoff}(1999)}]{kadanoff:99}%
  \BibitemOpen
  \bibfield  {author} {\bibinfo {author} {\bibfnamefont {L.~P.}\ \bibnamefont
  {Kadanoff}},\ }\href@noop {} {\bibfield  {journal} {\bibinfo  {journal} {Rev.
  Mod. Phys.}\ }\textbf {\bibinfo {volume} {71}},\ \bibinfo {pages} {435}
  (\bibinfo {year} {1999})}\BibitemShut {NoStop}%
\bibitem [{\citenamefont {Soto}\ \emph {et~al.}(1999)\citenamefont {Soto},
  \citenamefont {Mareschal},\ and\ \citenamefont {Risso}}]{soto:99}%
  \BibitemOpen
  \bibfield  {author} {\bibinfo {author} {\bibfnamefont {R.}~\bibnamefont
  {Soto}}, \bibinfo {author} {\bibfnamefont {M.}~\bibnamefont {Mareschal}}, \
  and\ \bibinfo {author} {\bibfnamefont {D.}~\bibnamefont {Risso}},\
  }\href@noop {} {\bibfield  {journal} {\bibinfo  {journal} {Phys. Rev. Lett.}\
  }\textbf {\bibinfo {volume} {83}},\ \bibinfo {pages} {5003} (\bibinfo {year}
  {1999})}\BibitemShut {NoStop}%
\bibitem [{\citenamefont {Dufty}(2007)}]{dufty:07}%
  \BibitemOpen
  \bibfield  {author} {\bibinfo {author} {\bibfnamefont {J.~W.}\ \bibnamefont
  {Dufty}},\ }\href@noop {} {\bibfield  {journal} {\bibinfo  {journal} {J.
  Phys. Chem. C}\ }\textbf {\bibinfo {volume} {111}},\ \bibinfo {pages} {15605}
  (\bibinfo {year} {2007})}\BibitemShut {NoStop}%
\bibitem [{\citenamefont {Candela}\ and\ \citenamefont
  {Walsworth}(2007)}]{candela:07}%
  \BibitemOpen
  \bibfield  {author} {\bibinfo {author} {\bibfnamefont {D.}~\bibnamefont
  {Candela}}\ and\ \bibinfo {author} {\bibfnamefont {R.~L.}\ \bibnamefont
  {Walsworth}},\ }\href@noop {} {\bibfield  {journal} {\bibinfo  {journal} {Am.
  J. Phys.}\ }\textbf {\bibinfo {volume} {75}},\ \bibinfo {pages} {754}
  (\bibinfo {year} {2007})}\BibitemShut {NoStop}%
\bibitem [{\citenamefont {Brilliantov}\ \emph {et~al.}(2007)\citenamefont
  {Brilliantov}, \citenamefont {P\"oschel}, \citenamefont {Kranz},\ and\
  \citenamefont {Zippelius}}]{brilliantov:07}%
  \BibitemOpen
  \bibfield  {author} {\bibinfo {author} {\bibfnamefont {N.~V.}\ \bibnamefont
  {Brilliantov}}, \bibinfo {author} {\bibfnamefont {T.}~\bibnamefont
  {P\"oschel}}, \bibinfo {author} {\bibfnamefont {W.~T.}\ \bibnamefont
  {Kranz}}, \ and\ \bibinfo {author} {\bibfnamefont {A.}~\bibnamefont
  {Zippelius}},\ }\href {\doibase 10.1103/PhysRevLett.98.128001} {\bibfield
  {journal} {\bibinfo  {journal} {Phys. Rev. Lett.}\ }\textbf {\bibinfo
  {volume} {98}},\ \bibinfo {pages} {128001} (\bibinfo {year}
  {2007})}\BibitemShut {NoStop}%
\bibitem [{\citenamefont {Kranz}\ \emph {et~al.}(2009)\citenamefont {Kranz},
  \citenamefont {Brilliantov}, \citenamefont {P\"oschel},\ and\ \citenamefont
  {Zippelius}}]{kranz:09}%
  \BibitemOpen
  \bibfield  {author} {\bibinfo {author} {\bibfnamefont {W.}~\bibnamefont
  {Kranz}}, \bibinfo {author} {\bibfnamefont {N.}~\bibnamefont {Brilliantov}},
  \bibinfo {author} {\bibfnamefont {T.}~\bibnamefont {P\"oschel}}, \ and\
  \bibinfo {author} {\bibfnamefont {A.}~\bibnamefont {Zippelius}},\ }\href
  {\doibase 10.1140/epjst/e2010-01196-0} {\bibfield  {journal} {\bibinfo
  {journal} {Eur. Phys. J. Special Topics}\ }\textbf {\bibinfo {volume}
  {179}},\ \bibinfo {pages} {91} (\bibinfo {year} {2009})}\BibitemShut
  {NoStop}%
\bibitem [{\citenamefont {Gayen}\ and\ \citenamefont {Alam}(2011)}]{gayen:11}%
  \BibitemOpen
  \bibfield  {author} {\bibinfo {author} {\bibfnamefont {B.}~\bibnamefont
  {Gayen}}\ and\ \bibinfo {author} {\bibfnamefont {M.}~\bibnamefont {Alam}},\
  }\href@noop {} {\bibfield  {journal} {\bibinfo  {journal} {Phys. Rev. E}\
  }\textbf {\bibinfo {volume} {84}},\ \bibinfo {pages} {021304} (\bibinfo
  {year} {2011})}\BibitemShut {NoStop}%
\bibitem [{\citenamefont {P\"oschel}\ \emph {et~al.}(2002)\citenamefont
  {P\"oschel}, \citenamefont {Brilliantov},\ and\ \citenamefont
  {Schwager}}]{poeschel:02}%
  \BibitemOpen
  \bibfield  {author} {\bibinfo {author} {\bibfnamefont {T.}~\bibnamefont
  {P\"oschel}}, \bibinfo {author} {\bibfnamefont {N.~V.}\ \bibnamefont
  {Brilliantov}}, \ and\ \bibinfo {author} {\bibfnamefont {T.}~\bibnamefont
  {Schwager}},\ }\href@noop {} {\bibfield  {journal} {\bibinfo  {journal} {Int.
  J. Mod. Phys. C}\ }\textbf {\bibinfo {volume} {13}},\ \bibinfo {pages} {1263}
  (\bibinfo {year} {2002})}\BibitemShut {NoStop}%
\bibitem [{\citenamefont {Shaw}\ \emph {et~al.}(2007)\citenamefont {Shaw},
  \citenamefont {Packard}, \citenamefont {Schr\"oter},\ and\ \citenamefont
  {Swinney}}]{shaw:07}%
  \BibitemOpen
  \bibfield  {author} {\bibinfo {author} {\bibfnamefont {R.~S.}\ \bibnamefont
  {Shaw}}, \bibinfo {author} {\bibfnamefont {N.}~\bibnamefont {Packard}},
  \bibinfo {author} {\bibfnamefont {M.}~\bibnamefont {Schr\"oter}}, \ and\
  \bibinfo {author} {\bibfnamefont {H.~L.}\ \bibnamefont {Swinney}},\
  }\href@noop {} {\bibfield  {journal} {\bibinfo  {journal} {PNAS}\ }\textbf
  {\bibinfo {volume} {104}},\ \bibinfo {pages} {9580} (\bibinfo {year}
  {2007})}\BibitemShut {NoStop}%
\bibitem [{\citenamefont {Hsiau}\ and\ \citenamefont {Hunt}(1996)}]{hsiau:96}%
  \BibitemOpen
  \bibfield  {author} {\bibinfo {author} {\bibfnamefont {S.~S.}\ \bibnamefont
  {Hsiau}}\ and\ \bibinfo {author} {\bibfnamefont {M.~L.}\ \bibnamefont
  {Hunt}},\ }\href@noop {} {\bibfield  {journal} {\bibinfo  {journal} {Acta
  Mech.}\ }\textbf {\bibinfo {volume} {114}},\ \bibinfo {pages} {121} (\bibinfo
  {year} {1996})}\BibitemShut {NoStop}%
\bibitem [{\citenamefont {Jenkins}\ and\ \citenamefont
  {Yoon}(2002)}]{jenkins:02}%
  \BibitemOpen
  \bibfield  {author} {\bibinfo {author} {\bibfnamefont {J.~T.}\ \bibnamefont
  {Jenkins}}\ and\ \bibinfo {author} {\bibfnamefont {D.~K.}\ \bibnamefont
  {Yoon}},\ }\href@noop {} {\bibfield  {journal} {\bibinfo  {journal} {Phys.
  Rev. Lett.}\ }\textbf {\bibinfo {volume} {88}},\ \bibinfo {pages} {194301}
  (\bibinfo {year} {2002})}\BibitemShut {NoStop}%
\bibitem [{\citenamefont {Schr\"oter}\ \emph {et~al.}(2006)\citenamefont
  {Schr\"oter}, \citenamefont {Ulrich}, \citenamefont {Kreft}, \citenamefont
  {Swift},\ and\ \citenamefont {Swinney}}]{schroeter:06}%
  \BibitemOpen
  \bibfield  {author} {\bibinfo {author} {\bibfnamefont {M.}~\bibnamefont
  {Schr\"oter}}, \bibinfo {author} {\bibfnamefont {S.}~\bibnamefont {Ulrich}},
  \bibinfo {author} {\bibfnamefont {J.}~\bibnamefont {Kreft}}, \bibinfo
  {author} {\bibfnamefont {J.~B.}\ \bibnamefont {Swift}}, \ and\ \bibinfo
  {author} {\bibfnamefont {H.~L.}\ \bibnamefont {Swinney}},\ }\href@noop {}
  {\bibfield  {journal} {\bibinfo  {journal} {Phys. Rev. E}\ }\textbf {\bibinfo
  {volume} {74}},\ \bibinfo {pages} {011307} (\bibinfo {year}
  {2006})}\BibitemShut {NoStop}%
\bibitem [{\citenamefont {Garz\'o}(2006)}]{garzo:06}%
  \BibitemOpen
  \bibfield  {author} {\bibinfo {author} {\bibfnamefont {V.}~\bibnamefont
  {Garz\'o}},\ }\href@noop {} {\bibfield  {journal} {\bibinfo  {journal}
  {Europhys. Lett.}\ }\textbf {\bibinfo {volume} {75}},\ \bibinfo {pages} {521}
  (\bibinfo {year} {2006})}\BibitemShut {NoStop}%
\bibitem [{\citenamefont {Garz\'o}(2011)}]{garzo:11}%
  \BibitemOpen
  \bibfield  {author} {\bibinfo {author} {\bibfnamefont {V.}~\bibnamefont
  {Garz\'o}},\ }\href@noop {} {\bibfield  {journal} {\bibinfo  {journal} {New
  J. Phys.}\ }\textbf {\bibinfo {volume} {13}},\ \bibinfo {pages} {055020}
  (\bibinfo {year} {2011})}\BibitemShut {NoStop}%
\bibitem [{\citenamefont {Brey}\ \emph {et~al.}(2011)\citenamefont {Brey},
  \citenamefont {Khalil},\ and\ \citenamefont {Dufty}}]{brey:11}%
  \BibitemOpen
  \bibfield  {author} {\bibinfo {author} {\bibfnamefont {J.~J.}\ \bibnamefont
  {Brey}}, \bibinfo {author} {\bibfnamefont {N.}~\bibnamefont {Khalil}}, \ and\
  \bibinfo {author} {\bibfnamefont {J.~W.}\ \bibnamefont {Dufty}},\ }\href@noop
  {} {\bibfield  {journal} {\bibinfo  {journal} {N. J. Phys.}\ }\textbf
  {\bibinfo {volume} {13}},\ \bibinfo {pages} {055019} (\bibinfo {year}
  {2011})}\BibitemShut {NoStop}%
\bibitem [{\citenamefont {Garz\'o}(2019)}]{garzo:19}%
  \BibitemOpen
  \bibfield  {author} {\bibinfo {author} {\bibfnamefont {V.}~\bibnamefont
  {Garz\'o}},\ }\href@noop {} {\emph {\bibinfo {title} {Granular Gaseous
  Flows}}}\ (\bibinfo  {publisher} {Springer},\ \bibinfo {year}
  {2019})\BibitemShut {NoStop}%
\bibitem [{\citenamefont {Serero}\ \emph {et~al.}(2008)\citenamefont {Serero},
  \citenamefont {Goldenberg}, \citenamefont {Noskowicz},\ and\ \citenamefont
  {Goldhirsch}}]{serero:08}%
  \BibitemOpen
  \bibfield  {author} {\bibinfo {author} {\bibfnamefont {D.}~\bibnamefont
  {Serero}}, \bibinfo {author} {\bibfnamefont {C.}~\bibnamefont {Goldenberg}},
  \bibinfo {author} {\bibfnamefont {S.~H.}\ \bibnamefont {Noskowicz}}, \ and\
  \bibinfo {author} {\bibfnamefont {I.}~\bibnamefont {Goldhirsch}},\ }\href
  {\doibase 10.1016/j.powtec.2007.08.002} {\bibfield  {journal} {\bibinfo
  {journal} {Powder Tech.}\ }\textbf {\bibinfo {volume} {182}},\ \bibinfo
  {pages} {257} (\bibinfo {year} {2008})}\BibitemShut {NoStop}%
\bibitem [{\citenamefont {Goldshtein}\ and\ \citenamefont
  {Shapiro}(1995)}]{Goldshtein1995}%
  \BibitemOpen
  \bibfield  {author} {\bibinfo {author} {\bibfnamefont {A.}~\bibnamefont
  {Goldshtein}}\ and\ \bibinfo {author} {\bibfnamefont {M.}~\bibnamefont
  {Shapiro}},\ }\href@noop {} {\bibfield  {journal} {\bibinfo  {journal} {J.
  Fluid Mech.}\ }\textbf {\bibinfo {volume} {282}},\ \bibinfo {pages} {75}
  (\bibinfo {year} {1995})}\BibitemShut {NoStop}%
\bibitem [{\citenamefont {Feitosa}\ and\ \citenamefont
  {Menon}(2002)}]{feitosa:02}%
  \BibitemOpen
  \bibfield  {author} {\bibinfo {author} {\bibfnamefont {K.}~\bibnamefont
  {Feitosa}}\ and\ \bibinfo {author} {\bibfnamefont {N.}~\bibnamefont
  {Menon}},\ }\href {\doibase 10.1103/PhysRevLett.88.198301} {\bibfield
  {journal} {\bibinfo  {journal} {Phys. Rev. Lett.}\ }\textbf {\bibinfo
  {volume} {88}},\ \bibinfo {pages} {198301} (\bibinfo {year}
  {2002})}\BibitemShut {NoStop}%
\bibitem [{\citenamefont {Galvin}\ \emph {et~al.}(2005)\citenamefont {Galvin},
  \citenamefont {Dahl},\ and\ \citenamefont {Hrenya}}]{galvin:05}%
  \BibitemOpen
  \bibfield  {author} {\bibinfo {author} {\bibfnamefont {J.~E.}\ \bibnamefont
  {Galvin}}, \bibinfo {author} {\bibfnamefont {S.~R.}\ \bibnamefont {Dahl}}, \
  and\ \bibinfo {author} {\bibfnamefont {C.~M.}\ \bibnamefont {Hrenya}},\
  }\href@noop {} {\bibfield  {journal} {\bibinfo  {journal} {J. Fluid Mech.}\
  }\textbf {\bibinfo {volume} {528}},\ \bibinfo {pages} {207} (\bibinfo {year}
  {2005})}\BibitemShut {NoStop}%
\bibitem [{\citenamefont {Harth}\ \emph {et~al.}(2013)\citenamefont {Harth},
  \citenamefont {Kornek}, \citenamefont {Trittel}, \citenamefont {Strachauer},
  \citenamefont {H\"ome}, \citenamefont {Will},\ and\ \citenamefont
  {Stannarius}}]{harth:13}%
  \BibitemOpen
  \bibfield  {author} {\bibinfo {author} {\bibfnamefont {K.}~\bibnamefont
  {Harth}}, \bibinfo {author} {\bibfnamefont {U.}~\bibnamefont {Kornek}},
  \bibinfo {author} {\bibfnamefont {T.}~\bibnamefont {Trittel}}, \bibinfo
  {author} {\bibfnamefont {U.}~\bibnamefont {Strachauer}}, \bibinfo {author}
  {\bibfnamefont {S.}~\bibnamefont {H\"ome}}, \bibinfo {author} {\bibfnamefont
  {K.}~\bibnamefont {Will}}, \ and\ \bibinfo {author} {\bibfnamefont
  {R.}~\bibnamefont {Stannarius}},\ }\href {\doibase
  10.1103/PhysRevLett.110.144102} {\bibfield  {journal} {\bibinfo  {journal}
  {Phys. Rev. Lett.}\ }\textbf {\bibinfo {volume} {110}},\ \bibinfo {pages}
  {144102} (\bibinfo {year} {2013})}\BibitemShut {NoStop}%
\bibitem [{\citenamefont {Goldhirsch}\ and\ \citenamefont
  {Zanetti}(1993)}]{goldhirsch:93}%
  \BibitemOpen
  \bibfield  {author} {\bibinfo {author} {\bibfnamefont {I.}~\bibnamefont
  {Goldhirsch}}\ and\ \bibinfo {author} {\bibfnamefont {G.}~\bibnamefont
  {Zanetti}},\ }\href@noop {} {\bibfield  {journal} {\bibinfo  {journal} {Phys.
  Rev. Lett.}\ }\textbf {\bibinfo {volume} {70}},\ \bibinfo {pages} {1619}
  (\bibinfo {year} {1993})}\BibitemShut {NoStop}%
\bibitem [{\citenamefont {Goldhirsch}\ \emph {et~al.}(1993)\citenamefont
  {Goldhirsch}, \citenamefont {Tan},\ and\ \citenamefont
  {Zanetti}}]{Goldhirsch1993_MDStudy}%
  \BibitemOpen
  \bibfield  {author} {\bibinfo {author} {\bibfnamefont {I.}~\bibnamefont
  {Goldhirsch}}, \bibinfo {author} {\bibfnamefont {M.-L.}\ \bibnamefont {Tan}},
  \ and\ \bibinfo {author} {\bibfnamefont {G.}~\bibnamefont {Zanetti}},\
  }\href@noop {} {\bibfield  {journal} {\bibinfo  {journal} {J. Sci. Comput.}\
  }\textbf {\bibinfo {volume} {8}},\ \bibinfo {pages} {1} (\bibinfo {year}
  {1993})}\BibitemShut {NoStop}%
\bibitem [{\citenamefont {Kudrolli}\ \emph {et~al.}(1997)\citenamefont
  {Kudrolli}, \citenamefont {Wolpert},\ and\ \citenamefont
  {Gollub}}]{kudrolli:97}%
  \BibitemOpen
  \bibfield  {author} {\bibinfo {author} {\bibfnamefont {A.}~\bibnamefont
  {Kudrolli}}, \bibinfo {author} {\bibfnamefont {M.}~\bibnamefont {Wolpert}}, \
  and\ \bibinfo {author} {\bibfnamefont {J.~P.}\ \bibnamefont {Gollub}},\
  }\href {\doibase 10.1103/PhysRevLett.78.1383} {\bibfield  {journal} {\bibinfo
   {journal} {Phys. Rev. Lett.}\ }\textbf {\bibinfo {volume} {78}},\ \bibinfo
  {pages} {1383} (\bibinfo {year} {1997})}\BibitemShut {NoStop}%
\bibitem [{\citenamefont {Falcon}\ \emph {et~al.}(1999)\citenamefont {Falcon},
  \citenamefont {Wunenburger}, \citenamefont {Evesque}, \citenamefont {Fauve},
  \citenamefont {Chabot}, \citenamefont {Garrabos},\ and\ \citenamefont
  {Beysens}}]{Falcon1999}%
  \BibitemOpen
  \bibfield  {author} {\bibinfo {author} {\bibfnamefont {E.}~\bibnamefont
  {Falcon}}, \bibinfo {author} {\bibfnamefont {R.}~\bibnamefont {Wunenburger}},
  \bibinfo {author} {\bibfnamefont {P.}~\bibnamefont {Evesque}}, \bibinfo
  {author} {\bibfnamefont {S.}~\bibnamefont {Fauve}}, \bibinfo {author}
  {\bibfnamefont {C.}~\bibnamefont {Chabot}}, \bibinfo {author} {\bibfnamefont
  {Y.}~\bibnamefont {Garrabos}}, \ and\ \bibinfo {author} {\bibfnamefont
  {D.}~\bibnamefont {Beysens}},\ }\href@noop {} {\bibfield  {journal} {\bibinfo
   {journal} {Phys. Rev. Lett.}\ }\textbf {\bibinfo {volume} {83}},\ \bibinfo
  {pages} {440} (\bibinfo {year} {1999})}\BibitemShut {NoStop}%
\bibitem [{\citenamefont {Opsomer}\ \emph {et~al.}(2011)\citenamefont
  {Opsomer}, \citenamefont {Ludewig},\ and\ \citenamefont
  {Vandewalle}}]{opsomer:11}%
  \BibitemOpen
  \bibfield  {author} {\bibinfo {author} {\bibfnamefont {E.}~\bibnamefont
  {Opsomer}}, \bibinfo {author} {\bibfnamefont {F.}~\bibnamefont {Ludewig}}, \
  and\ \bibinfo {author} {\bibfnamefont {N.}~\bibnamefont {Vandewalle}},\
  }\href {\doibase 10.1103/PhysRevE.84.051306} {\bibfield  {journal} {\bibinfo
  {journal} {Phys. Rev. E}\ }\textbf {\bibinfo {volume} {84}},\ \bibinfo
  {pages} {051306} (\bibinfo {year} {2011})}\BibitemShut {NoStop}%
\bibitem [{\citenamefont {Noirhomme}\ \emph {et~al.}(2018)\citenamefont
  {Noirhomme}, \citenamefont {Cazaubiel}, \citenamefont {Darras}, \citenamefont
  {Falcon}, \citenamefont {Fischer}, \citenamefont {Garrabos}, \citenamefont
  {Lecoutre-Chabot}, \citenamefont {Merminod}, \citenamefont {Opsomer},
  \citenamefont {Palencia}, \citenamefont {Schockmel}, \citenamefont
  {Stannarius},\ and\ \citenamefont {Vandewalle}}]{noirhomme:18}%
  \BibitemOpen
  \bibfield  {author} {\bibinfo {author} {\bibfnamefont {M.}~\bibnamefont
  {Noirhomme}}, \bibinfo {author} {\bibfnamefont {A.}~\bibnamefont
  {Cazaubiel}}, \bibinfo {author} {\bibfnamefont {A.}~\bibnamefont {Darras}},
  \bibinfo {author} {\bibfnamefont {E.}~\bibnamefont {Falcon}}, \bibinfo
  {author} {\bibfnamefont {D.}~\bibnamefont {Fischer}}, \bibinfo {author}
  {\bibfnamefont {Y.}~\bibnamefont {Garrabos}}, \bibinfo {author}
  {\bibfnamefont {C.}~\bibnamefont {Lecoutre-Chabot}}, \bibinfo {author}
  {\bibfnamefont {S.}~\bibnamefont {Merminod}}, \bibinfo {author}
  {\bibfnamefont {E.}~\bibnamefont {Opsomer}}, \bibinfo {author} {\bibfnamefont
  {F.}~\bibnamefont {Palencia}}, \bibinfo {author} {\bibfnamefont
  {J.}~\bibnamefont {Schockmel}}, \bibinfo {author} {\bibfnamefont
  {R.}~\bibnamefont {Stannarius}}, \ and\ \bibinfo {author} {\bibfnamefont
  {N.}~\bibnamefont {Vandewalle}},\ }\href {\doibase
  10.1209/0295-5075/123/14003} {\bibfield  {journal} {\bibinfo  {journal}
  {Europhys. Lett.}\ }\textbf {\bibinfo {volume} {123}},\ \bibinfo {pages}
  {14003} (\bibinfo {year} {2018})}\BibitemShut {NoStop}%
\bibitem [{\citenamefont {Mitrano}\ \emph {et~al.}(2012)\citenamefont
  {Mitrano}, \citenamefont {Garz\'o}, \citenamefont {Hilger}, \citenamefont
  {Ewasko},\ and\ \citenamefont {Hrenya}}]{mitrano:12}%
  \BibitemOpen
  \bibfield  {author} {\bibinfo {author} {\bibfnamefont {P.~P.}\ \bibnamefont
  {Mitrano}}, \bibinfo {author} {\bibfnamefont {V.}~\bibnamefont {Garz\'o}},
  \bibinfo {author} {\bibfnamefont {A.~M.}\ \bibnamefont {Hilger}}, \bibinfo
  {author} {\bibfnamefont {C.~J.}\ \bibnamefont {Ewasko}}, \ and\ \bibinfo
  {author} {\bibfnamefont {C.~M.}\ \bibnamefont {Hrenya}},\ }\href {\doibase
  10.1103/PhysRevE.85.041303} {\bibfield  {journal} {\bibinfo  {journal} {Phys.
  Rev. E}\ }\textbf {\bibinfo {volume} {85}},\ \bibinfo {pages} {041303}
  (\bibinfo {year} {2012})}\BibitemShut {NoStop}%
\bibitem [{\citenamefont {Maa\ss{}}\ \emph {et~al.}(2008)\citenamefont
  {Maa\ss{}}, \citenamefont {Isert}, \citenamefont {Maret},\ and\ \citenamefont
  {Aegerter}}]{Maass2008}%
  \BibitemOpen
  \bibfield  {author} {\bibinfo {author} {\bibfnamefont {C.}~\bibnamefont
  {Maa\ss{}}}, \bibinfo {author} {\bibfnamefont {N.}~\bibnamefont {Isert}},
  \bibinfo {author} {\bibfnamefont {G.}~\bibnamefont {Maret}}, \ and\ \bibinfo
  {author} {\bibfnamefont {C.~M.}\ \bibnamefont {Aegerter}},\ }\href@noop {}
  {\bibfield  {journal} {\bibinfo  {journal} {Phys. Rev. Lett.}\ }\textbf
  {\bibinfo {volume} {100}},\ \bibinfo {pages} {248001} (\bibinfo {year}
  {2008})}\BibitemShut {NoStop}%
\bibitem [{\citenamefont {Hummel}\ \emph {et~al.}(2016)\citenamefont {Hummel},
  \citenamefont {Clewett},\ and\ \citenamefont {Mazza}}]{hummel:16}%
  \BibitemOpen
  \bibfield  {author} {\bibinfo {author} {\bibfnamefont {M.}~\bibnamefont
  {Hummel}}, \bibinfo {author} {\bibfnamefont {J.~P.~D.}\ \bibnamefont
  {Clewett}}, \ and\ \bibinfo {author} {\bibfnamefont {M.~G.}\ \bibnamefont
  {Mazza}},\ }\href {\doibase 10.1209/0295-5075/114/10002} {\bibfield
  {journal} {\bibinfo  {journal} {Europhys. Lett.}\ }\textbf {\bibinfo {volume}
  {114}},\ \bibinfo {pages} {10002} (\bibinfo {year} {2016})}\BibitemShut
  {NoStop}%
\bibitem [{\citenamefont {Opsomer}\ \emph {et~al.}(2017)\citenamefont
  {Opsomer}, \citenamefont {Noirhomme}, \citenamefont {Vandewalle},
  \citenamefont {Falcon},\ and\ \citenamefont {Merminod}}]{opsomer:17}%
  \BibitemOpen
  \bibfield  {author} {\bibinfo {author} {\bibfnamefont {E.}~\bibnamefont
  {Opsomer}}, \bibinfo {author} {\bibfnamefont {M.}~\bibnamefont {Noirhomme}},
  \bibinfo {author} {\bibfnamefont {N.}~\bibnamefont {Vandewalle}}, \bibinfo
  {author} {\bibfnamefont {E.}~\bibnamefont {Falcon}}, \ and\ \bibinfo {author}
  {\bibfnamefont {S.}~\bibnamefont {Merminod}},\ }\href {\doibase
  10.1038/s41526-016-0009-1} {\bibfield  {journal} {\bibinfo  {journal} {NPJ
  Microgravity}\ }\textbf {\bibinfo {volume} {3}},\ \bibinfo {pages} {1}
  (\bibinfo {year} {2017})}\BibitemShut {NoStop}%
\bibitem [{\citenamefont {Nichol}\ and\ \citenamefont
  {Daniels}(2012)}]{nichol:12}%
  \BibitemOpen
  \bibfield  {author} {\bibinfo {author} {\bibfnamefont {K.}~\bibnamefont
  {Nichol}}\ and\ \bibinfo {author} {\bibfnamefont {K.~E.}\ \bibnamefont
  {Daniels}},\ }\href {\doibase 10.1103/PhysRevLett.108.018001} {\bibfield
  {journal} {\bibinfo  {journal} {Phys. Rev. Lett.}\ }\textbf {\bibinfo
  {volume} {108}},\ \bibinfo {pages} {018001} (\bibinfo {year}
  {2012})}\BibitemShut {NoStop}%
\bibitem [{\citenamefont {Chapman}\ and\ \citenamefont
  {Cowling}(1960)}]{Chapman1960}%
  \BibitemOpen
  \bibfield  {author} {\bibinfo {author} {\bibfnamefont {S.}~\bibnamefont
  {Chapman}}\ and\ \bibinfo {author} {\bibfnamefont {T.~G.}\ \bibnamefont
  {Cowling}},\ }\href@noop {} {\emph {\bibinfo {title} {The mathematical theory
  of nonuniform gases}}}\ (\bibinfo  {publisher} {Cambridge University Press},\
  \bibinfo {address} {London},\ \bibinfo {year} {1960})\BibitemShut {NoStop}%
\bibitem [{\citenamefont {Van~Noije}\ and\ \citenamefont
  {Ernst}(1998)}]{vanNoije1998}%
  \BibitemOpen
  \bibfield  {author} {\bibinfo {author} {\bibfnamefont {T.}~\bibnamefont
  {Van~Noije}}\ and\ \bibinfo {author} {\bibfnamefont {M.}~\bibnamefont
  {Ernst}},\ }\href@noop {} {\bibfield  {journal} {\bibinfo  {journal} {Granul.
  Matter}\ }\textbf {\bibinfo {volume} {1}},\ \bibinfo {pages} {57} (\bibinfo
  {year} {1998})}\BibitemShut {NoStop}%
\bibitem [{\citenamefont {Esipov}\ and\ \citenamefont
  {P\"oschel}(1997)}]{esipov:97}%
  \BibitemOpen
  \bibfield  {author} {\bibinfo {author} {\bibfnamefont {S.~E.}\ \bibnamefont
  {Esipov}}\ and\ \bibinfo {author} {\bibfnamefont {T.}~\bibnamefont
  {P\"oschel}},\ }\href@noop {} {\bibfield  {journal} {\bibinfo  {journal} {J.
  Stat. Phys.}\ }\textbf {\bibinfo {volume} {86}},\ \bibinfo {pages} {1385}
  (\bibinfo {year} {1997})}\BibitemShut {NoStop}%
\bibitem [{\citenamefont {Puglisi}\ \emph {et~al.}(1999)\citenamefont
  {Puglisi}, \citenamefont {Loreto}, \citenamefont {Marini Bettolo~Marconi},\
  and\ \citenamefont {Vulpiani}}]{puglisi:99}%
  \BibitemOpen
  \bibfield  {author} {\bibinfo {author} {\bibfnamefont {A.}~\bibnamefont
  {Puglisi}}, \bibinfo {author} {\bibfnamefont {V.}~\bibnamefont {Loreto}},
  \bibinfo {author} {\bibfnamefont {U.}~\bibnamefont {Marini Bettolo~Marconi}},
  \ and\ \bibinfo {author} {\bibfnamefont {A.}~\bibnamefont {Vulpiani}},\
  }\href {\doibase 10.1103/PhysRevE.59.5582} {\bibfield  {journal} {\bibinfo
  {journal} {Phys. Rev. E}\ }\textbf {\bibinfo {volume} {59}},\ \bibinfo
  {pages} {5582} (\bibinfo {year} {1999})}\BibitemShut {NoStop}%
\bibitem [{\citenamefont {Goldhirsch}\ \emph {et~al.}(2003)\citenamefont
  {Goldhirsch}, \citenamefont {Noskowicz},\ and\ \citenamefont
  {Bar-Lev}}]{goldhirsch:03}%
  \BibitemOpen
  \bibfield  {author} {\bibinfo {author} {\bibfnamefont {I.}~\bibnamefont
  {Goldhirsch}}, \bibinfo {author} {\bibfnamefont {S.~H.}\ \bibnamefont
  {Noskowicz}}, \ and\ \bibinfo {author} {\bibfnamefont {O.}~\bibnamefont
  {Bar-Lev}},\ }in\ \href {\doibase 10.1007/978-3-540-39843-1_2} {\emph
  {\bibinfo {booktitle} {Granular {Gas} {Dynamics}}}},\ \bibinfo {series and
  number} {Lecture {Notes} in {Physics}},\ \bibinfo {editor} {edited by\
  \bibinfo {editor} {\bibfnamefont {T.}~\bibnamefont {P\"oschel}}\ and\
  \bibinfo {editor} {\bibfnamefont {N.}~\bibnamefont {Brilliantov}}}\ (\bibinfo
   {publisher} {Springer Berlin Heidelberg},\ \bibinfo {address} {Berlin,
  Heidelberg},\ \bibinfo {year} {2003})\ pp.\ \bibinfo {pages}
  {37--63}\BibitemShut {NoStop}%
\bibitem [{\citenamefont {P\"oschel}\ \emph {et~al.}(2006)\citenamefont
  {P\"oschel}, \citenamefont {Brilliantov},\ and\ \citenamefont
  {Formella}}]{poeschel:06}%
  \BibitemOpen
  \bibfield  {author} {\bibinfo {author} {\bibfnamefont {T.}~\bibnamefont
  {P\"oschel}}, \bibinfo {author} {\bibfnamefont {N.~V.}\ \bibnamefont
  {Brilliantov}}, \ and\ \bibinfo {author} {\bibfnamefont {A.}~\bibnamefont
  {Formella}},\ }\href {\doibase 10.1103/PhysRevE.74.041302} {\bibfield
  {journal} {\bibinfo  {journal} {Phys. Rev. E}\ }\textbf {\bibinfo {volume}
  {74}},\ \bibinfo {pages} {041302} (\bibinfo {year} {2006})}\BibitemShut
  {NoStop}%
\bibitem [{\citenamefont {P\"oschel}\ \emph {et~al.}(2007)\citenamefont
  {P\"oschel}, \citenamefont {Brilliantov},\ and\ \citenamefont
  {Formella}}]{poeschel:07}%
  \BibitemOpen
  \bibfield  {author} {\bibinfo {author} {\bibfnamefont {T.}~\bibnamefont
  {P\"oschel}}, \bibinfo {author} {\bibfnamefont {N.~V.}\ \bibnamefont
  {Brilliantov}}, \ and\ \bibinfo {author} {\bibfnamefont {A.}~\bibnamefont
  {Formella}},\ }\href {\doibase 10.1142/S0129183107010966} {\bibfield
  {journal} {\bibinfo  {journal} {Int. J. Mod. Phys. C}\ }\textbf {\bibinfo
  {volume} {18}},\ \bibinfo {pages} {701} (\bibinfo {year} {2007})}\BibitemShut
  {NoStop}%
\bibitem [{\citenamefont {Haff}(1983)}]{Haff1983}%
  \BibitemOpen
  \bibfield  {author} {\bibinfo {author} {\bibfnamefont {P.~K.}\ \bibnamefont
  {Haff}},\ }\href@noop {} {\bibfield  {journal} {\bibinfo  {journal} {J. Fluid
  Mech.}\ }\textbf {\bibinfo {volume} {134}},\ \bibinfo {pages} {401} (\bibinfo
  {year} {1983})}\BibitemShut {NoStop}%
\bibitem [{sup()}]{supp_material}%
  \BibitemOpen
  \href {http://link.aps.org/supplemental/00.0000/PhysRevLett.000.000000}
  {}\bibinfo {howpublished} {See the Supplementary Material at
  \url{http://link.aps.org/supplemental/00.0000/PhysRevLett.000.000000} for the
  image processing methods, further elaboration of theories, additional details
  of statistical analysis, and the descriptions of the calibration experiment,
  which includes Refs.
  \cite{besseling:09,Brilliantov2003,dalitz:17,jaqaman:08,ouellette:06,poeschel:07}}\BibitemShut
  {NoStop}%
\bibitem [{\citenamefont {Bougie}\ \emph {et~al.}(2002)\citenamefont {Bougie},
  \citenamefont {Moon}, \citenamefont {Swift},\ and\ \citenamefont
  {Swinney}}]{bougie:02}%
  \BibitemOpen
  \bibfield  {author} {\bibinfo {author} {\bibfnamefont {J.}~\bibnamefont
  {Bougie}}, \bibinfo {author} {\bibfnamefont {S.~J.}\ \bibnamefont {Moon}},
  \bibinfo {author} {\bibfnamefont {J.~B.}\ \bibnamefont {Swift}}, \ and\
  \bibinfo {author} {\bibfnamefont {H.~L.}\ \bibnamefont {Swinney}},\
  }\href@noop {} {\bibfield  {journal} {\bibinfo  {journal} {Phys. Rev. E}\
  }\textbf {\bibinfo {volume} {66}},\ \bibinfo {pages} {051301} (\bibinfo
  {year} {2002})}\BibitemShut {NoStop}%
\bibitem [{\citenamefont {Meerson}\ \emph {et~al.}(2004)\citenamefont
  {Meerson}, \citenamefont {P\"oschel}, \citenamefont {Sasorov},\ and\
  \citenamefont {Schwager}}]{meerson:04}%
  \BibitemOpen
  \bibfield  {author} {\bibinfo {author} {\bibfnamefont {B.}~\bibnamefont
  {Meerson}}, \bibinfo {author} {\bibfnamefont {T.}~\bibnamefont {P\"oschel}},
  \bibinfo {author} {\bibfnamefont {P.~V.}\ \bibnamefont {Sasorov}}, \ and\
  \bibinfo {author} {\bibfnamefont {T.}~\bibnamefont {Schwager}},\ }\href
  {\doibase 10.1103/PhysRevE.69.021302} {\bibfield  {journal} {\bibinfo
  {journal} {Phys. Rev. E}\ }\textbf {\bibinfo {volume} {69}},\ \bibinfo
  {pages} {021302} (\bibinfo {year} {2004})}\BibitemShut {NoStop}%
\bibitem [{\citenamefont {Eshuis}\ \emph {et~al.}(2007)\citenamefont {Eshuis},
  \citenamefont {van~der Weele}, \citenamefont {van~der Meer}, \citenamefont
  {Bos},\ and\ \citenamefont {Lohse}}]{eshuis:07}%
  \BibitemOpen
  \bibfield  {author} {\bibinfo {author} {\bibfnamefont {P.}~\bibnamefont
  {Eshuis}}, \bibinfo {author} {\bibfnamefont {K.}~\bibnamefont {van~der
  Weele}}, \bibinfo {author} {\bibfnamefont {D.}~\bibnamefont {van~der Meer}},
  \bibinfo {author} {\bibfnamefont {R.}~\bibnamefont {Bos}}, \ and\ \bibinfo
  {author} {\bibfnamefont {D.}~\bibnamefont {Lohse}},\ }\href {\doibase
  doi:10.1063/1.2815745} {\bibfield  {journal} {\bibinfo  {journal} {Phys.
  Fluids}\ }\textbf {\bibinfo {volume} {19}},\ \bibinfo {pages} {123301}
  (\bibinfo {year} {2007})}\BibitemShut {NoStop}%
\bibitem [{\citenamefont {Sack}\ \emph {et~al.}(2013)\citenamefont {Sack},
  \citenamefont {Heckel}, \citenamefont {Kollmer}, \citenamefont {Zimber},\
  and\ \citenamefont {P\"oschel}}]{Sack2013}%
  \BibitemOpen
  \bibfield  {author} {\bibinfo {author} {\bibfnamefont {A.}~\bibnamefont
  {Sack}}, \bibinfo {author} {\bibfnamefont {M.}~\bibnamefont {Heckel}},
  \bibinfo {author} {\bibfnamefont {J.~E.}\ \bibnamefont {Kollmer}}, \bibinfo
  {author} {\bibfnamefont {F.}~\bibnamefont {Zimber}}, \ and\ \bibinfo {author}
  {\bibfnamefont {T.}~\bibnamefont {P\"oschel}},\ }\href@noop {} {\bibfield
  {journal} {\bibinfo  {journal} {Phys. Rev. Lett.}\ }\textbf {\bibinfo
  {volume} {111}},\ \bibinfo {pages} {018001} (\bibinfo {year}
  {2013})}\BibitemShut {NoStop}%
\bibitem [{\citenamefont {Olafsen}\ and\ \citenamefont
  {Urbach}(1998)}]{olafsen:98}%
  \BibitemOpen
  \bibfield  {author} {\bibinfo {author} {\bibfnamefont {J.~S.}\ \bibnamefont
  {Olafsen}}\ and\ \bibinfo {author} {\bibfnamefont {J.~S.}\ \bibnamefont
  {Urbach}},\ }\href {\doibase 10.1103/PhysRevLett.81.4369} {\bibfield
  {journal} {\bibinfo  {journal} {Phys. Rev. Lett.}\ }\textbf {\bibinfo
  {volume} {81}},\ \bibinfo {pages} {4369} (\bibinfo {year}
  {1998})}\BibitemShut {NoStop}%
\bibitem [{\citenamefont {Baxter}\ and\ \citenamefont
  {Olafsen}(2003)}]{baxter:03}%
  \BibitemOpen
  \bibfield  {author} {\bibinfo {author} {\bibfnamefont {G.~W.}\ \bibnamefont
  {Baxter}}\ and\ \bibinfo {author} {\bibfnamefont {J.~S.}\ \bibnamefont
  {Olafsen}},\ }\href {\doibase 10.1038/425680a} {\bibfield  {journal}
  {\bibinfo  {journal} {Nature}\ }\textbf {\bibinfo {volume} {425}},\ \bibinfo
  {pages} {680} (\bibinfo {year} {2003})}\BibitemShut {NoStop}%
\bibitem [{\citenamefont {Tatsumi}\ \emph {et~al.}(2009)\citenamefont
  {Tatsumi}, \citenamefont {Murayama}, \citenamefont {Hayakawa},\ and\
  \citenamefont {Sano}}]{tatsumi:09}%
  \BibitemOpen
  \bibfield  {author} {\bibinfo {author} {\bibfnamefont {S.}~\bibnamefont
  {Tatsumi}}, \bibinfo {author} {\bibfnamefont {Y.}~\bibnamefont {Murayama}},
  \bibinfo {author} {\bibfnamefont {H.}~\bibnamefont {Hayakawa}}, \ and\
  \bibinfo {author} {\bibfnamefont {M.}~\bibnamefont {Sano}},\ }\href {\doibase
  10.1017/S002211200999231X} {\bibfield  {journal} {\bibinfo  {journal} {J.
  Fluid Mech.}\ }\textbf {\bibinfo {volume} {641}},\ \bibinfo {pages} {521}
  (\bibinfo {year} {2009})}\BibitemShut {NoStop}%
\bibitem [{\citenamefont {{P. M. Reis}}\ \emph {et~al.}(2006)\citenamefont {{P.
  M. Reis}}, \citenamefont {{R. A. Ingale}},\ and\ \citenamefont {{M. D.
  Shattuck}}}]{reis:06}%
  \BibitemOpen
  \bibfield  {author} {\bibinfo {author} {\bibnamefont {{P. M. Reis}}},
  \bibinfo {author} {\bibnamefont {{R. A. Ingale}}}, \ and\ \bibinfo {author}
  {\bibnamefont {{M. D. Shattuck}}},\ }\href {\doibase
  10.1103/PhysRevLett.96.258001} {\bibfield  {journal} {\bibinfo  {journal}
  {Phys. Rev. Lett.}\ }\textbf {\bibinfo {volume} {96}},\ \bibinfo {pages}
  {258001} (\bibinfo {year} {2006})}\BibitemShut {NoStop}%
\bibitem [{\citenamefont {Deseigne}\ \emph {et~al.}(2010)\citenamefont
  {Deseigne}, \citenamefont {Dauchot},\ and\ \citenamefont
  {Chat\'e}}]{deseigne:10}%
  \BibitemOpen
  \bibfield  {author} {\bibinfo {author} {\bibfnamefont {J.}~\bibnamefont
  {Deseigne}}, \bibinfo {author} {\bibfnamefont {O.}~\bibnamefont {Dauchot}}, \
  and\ \bibinfo {author} {\bibfnamefont {H.}~\bibnamefont {Chat\'e}},\ }\href
  {\doibase 10.1103/PhysRevLett.105.098001} {\bibfield  {journal} {\bibinfo
  {journal} {Phys. Rev. Lett.}\ }\textbf {\bibinfo {volume} {105}},\ \bibinfo
  {pages} {098001} (\bibinfo {year} {2010})}\BibitemShut {NoStop}%
\bibitem [{\citenamefont {Schmick}\ and\ \citenamefont
  {Markus}(2008)}]{schmick:08}%
  \BibitemOpen
  \bibfield  {author} {\bibinfo {author} {\bibfnamefont {M.}~\bibnamefont
  {Schmick}}\ and\ \bibinfo {author} {\bibfnamefont {M.}~\bibnamefont
  {Markus}},\ }\href {\doibase 10.1103/PhysRevE.78.010302} {\bibfield
  {journal} {\bibinfo  {journal} {Phys. Rev. E}\ }\textbf {\bibinfo {volume}
  {78}},\ \bibinfo {pages} {010302} (\bibinfo {year} {2008})}\BibitemShut
  {NoStop}%
\bibitem [{\citenamefont {Falcon}\ \emph {et~al.}(2013)\citenamefont {Falcon},
  \citenamefont {Bacri},\ and\ \citenamefont {Laroche}}]{Falcon2013}%
  \BibitemOpen
  \bibfield  {author} {\bibinfo {author} {\bibfnamefont {E.}~\bibnamefont
  {Falcon}}, \bibinfo {author} {\bibfnamefont {J.-C.}\ \bibnamefont {Bacri}}, \
  and\ \bibinfo {author} {\bibfnamefont {C.}~\bibnamefont {Laroche}},\
  }\href@noop {} {\bibfield  {journal} {\bibinfo  {journal} {Europhys. Lett.}\
  }\textbf {\bibinfo {volume} {103}},\ \bibinfo {pages} {64004} (\bibinfo
  {year} {2013})}\BibitemShut {NoStop}%
\bibitem [{\citenamefont {Yu}\ \emph {et~al.}(2019)\citenamefont {Yu},
  \citenamefont {St\"ark}, \citenamefont {Blochberger}, \citenamefont {Kaplik},
  \citenamefont {Offermann}, \citenamefont {Tran}, \citenamefont {Adachi},\
  and\ \citenamefont {Sperl}}]{Yu2019}%
  \BibitemOpen
  \bibfield  {author} {\bibinfo {author} {\bibfnamefont {P.}~\bibnamefont
  {Yu}}, \bibinfo {author} {\bibfnamefont {E.}~\bibnamefont {St\"ark}},
  \bibinfo {author} {\bibfnamefont {G.}~\bibnamefont {Blochberger}}, \bibinfo
  {author} {\bibfnamefont {M.}~\bibnamefont {Kaplik}}, \bibinfo {author}
  {\bibfnamefont {M.}~\bibnamefont {Offermann}}, \bibinfo {author}
  {\bibfnamefont {D.}~\bibnamefont {Tran}}, \bibinfo {author} {\bibfnamefont
  {M.}~\bibnamefont {Adachi}}, \ and\ \bibinfo {author} {\bibfnamefont
  {M.}~\bibnamefont {Sperl}},\ }\href@noop {} {\bibfield  {journal} {\bibinfo
  {journal} {Rev. Sci. Instrum.}\ }\textbf {\bibinfo {volume} {90}},\ \bibinfo
  {pages} {054501} (\bibinfo {year} {2019})}\BibitemShut {NoStop}%
\bibitem [{\citenamefont {Adachi}\ \emph {et~al.}(2019)\citenamefont {Adachi},
  \citenamefont {Yu},\ and\ \citenamefont {Sperl}}]{Masato2019}%
  \BibitemOpen
  \bibfield  {author} {\bibinfo {author} {\bibfnamefont {M.}~\bibnamefont
  {Adachi}}, \bibinfo {author} {\bibfnamefont {P.}~\bibnamefont {Yu}}, \ and\
  \bibinfo {author} {\bibfnamefont {M.}~\bibnamefont {Sperl}},\ }\href@noop {}
  {\bibfield  {journal} {\bibinfo  {journal} {NPJ Microgravity}\ }\textbf
  {\bibinfo {volume} {19}},\ \bibinfo {pages} {1} (\bibinfo {year}
  {2019})}\BibitemShut {NoStop}%
\bibitem [{\citenamefont {Aranson}\ and\ \citenamefont
  {Olafsen}(2002)}]{aranson:02}%
  \BibitemOpen
  \bibfield  {author} {\bibinfo {author} {\bibfnamefont {I.~S.}\ \bibnamefont
  {Aranson}}\ and\ \bibinfo {author} {\bibfnamefont {J.~S.}\ \bibnamefont
  {Olafsen}},\ }\href {\doibase 10.1103/PhysRevE.66.061302} {\bibfield
  {journal} {\bibinfo  {journal} {Phys. Rev. E}\ }\textbf {\bibinfo {volume}
  {66}},\ \bibinfo {pages} {061302} (\bibinfo {year} {2002})}\BibitemShut
  {NoStop}%
\bibitem [{\citenamefont {Falcon}\ \emph {et~al.}(2006)\citenamefont {Falcon},
  \citenamefont {Auma\^itre}, \citenamefont {Evesque}, \citenamefont
  {Palencia}, \citenamefont {Lecoutre-Chabot}, \citenamefont {Fauve},
  \citenamefont {Beysens},\ and\ \citenamefont {Garrabos}}]{Falcon2006}%
  \BibitemOpen
  \bibfield  {author} {\bibinfo {author} {\bibfnamefont {E.}~\bibnamefont
  {Falcon}}, \bibinfo {author} {\bibfnamefont {S.}~\bibnamefont {Auma\^itre}},
  \bibinfo {author} {\bibfnamefont {P.}~\bibnamefont {Evesque}}, \bibinfo
  {author} {\bibfnamefont {F.}~\bibnamefont {Palencia}}, \bibinfo {author}
  {\bibfnamefont {C.}~\bibnamefont {Lecoutre-Chabot}}, \bibinfo {author}
  {\bibfnamefont {S.}~\bibnamefont {Fauve}}, \bibinfo {author} {\bibfnamefont
  {D.}~\bibnamefont {Beysens}}, \ and\ \bibinfo {author} {\bibfnamefont
  {Y.}~\bibnamefont {Garrabos}},\ }\href@noop {} {\bibfield  {journal}
  {\bibinfo  {journal} {Europhys. Lett.}\ }\textbf {\bibinfo {volume} {74}},\
  \bibinfo {pages} {830} (\bibinfo {year} {2006})}\BibitemShut {NoStop}%
\bibitem [{\citenamefont {Leconte}\ \emph {et~al.}(2006)\citenamefont
  {Leconte}, \citenamefont {Garrabos}, \citenamefont {Falcon}, \citenamefont
  {Lecoutre-Chabot}, \citenamefont {Palencia}, \citenamefont {\'Evesque},\ and\
  \citenamefont {Beysens}}]{leconte:06}%
  \BibitemOpen
  \bibfield  {author} {\bibinfo {author} {\bibfnamefont {M.}~\bibnamefont
  {Leconte}}, \bibinfo {author} {\bibfnamefont {Y.}~\bibnamefont {Garrabos}},
  \bibinfo {author} {\bibfnamefont {E.}~\bibnamefont {Falcon}}, \bibinfo
  {author} {\bibfnamefont {C.}~\bibnamefont {Lecoutre-Chabot}}, \bibinfo
  {author} {\bibfnamefont {F.}~\bibnamefont {Palencia}}, \bibinfo {author}
  {\bibfnamefont {P.}~\bibnamefont {\'Evesque}}, \ and\ \bibinfo {author}
  {\bibfnamefont {D.}~\bibnamefont {Beysens}},\ }\href {\doibase
  10.1088/1742-5468/2006/07/P07012} {\bibfield  {journal} {\bibinfo  {journal}
  {J. Stat. Mech.: Theory Exp.}\ ,\ \bibinfo {pages} {P07012}} (\bibinfo {year}
  {2006})}\BibitemShut {NoStop}%
\bibitem [{\citenamefont {Grasselli}\ \emph {et~al.}(2015)\citenamefont
  {Grasselli}, \citenamefont {Bossis},\ and\ \citenamefont
  {Morini}}]{grasselli:15}%
  \BibitemOpen
  \bibfield  {author} {\bibinfo {author} {\bibfnamefont {Y.}~\bibnamefont
  {Grasselli}}, \bibinfo {author} {\bibfnamefont {G.}~\bibnamefont {Bossis}}, \
  and\ \bibinfo {author} {\bibfnamefont {R.}~\bibnamefont {Morini}},\ }\href
  {\doibase 10.1140/epje/i2015-15008-5} {\bibfield  {journal} {\bibinfo
  {journal} {Euro. Phys. J. E}\ }\textbf {\bibinfo {volume} {38}},\ \bibinfo
  {pages} {8} (\bibinfo {year} {2015})}\BibitemShut {NoStop}%
\bibitem [{\citenamefont {Heißelmann}\ \emph {et~al.}(2010)\citenamefont
  {Heißelmann}, \citenamefont {Blum}, \citenamefont {Fraser},\ and\
  \citenamefont {Wolling}}]{heiselmann:10}%
  \BibitemOpen
  \bibfield  {author} {\bibinfo {author} {\bibfnamefont {D.}~\bibnamefont
  {Heißelmann}}, \bibinfo {author} {\bibfnamefont {J.}~\bibnamefont {Blum}},
  \bibinfo {author} {\bibfnamefont {H.~J.}\ \bibnamefont {Fraser}}, \ and\
  \bibinfo {author} {\bibfnamefont {K.}~\bibnamefont {Wolling}},\ }\href
  {\doibase 10.1016/j.icarus.2009.08.009} {\bibfield  {journal} {\bibinfo
  {journal} {Icarus}\ }\textbf {\bibinfo {volume} {206}},\ \bibinfo {pages}
  {424} (\bibinfo {year} {2010})}\BibitemShut {NoStop}%
\bibitem [{\citenamefont {Born}\ \emph {et~al.}(2017)\citenamefont {Born},
  \citenamefont {Schmitz},\ and\ \citenamefont {Sperl}}]{born:17}%
  \BibitemOpen
  \bibfield  {author} {\bibinfo {author} {\bibfnamefont {P.}~\bibnamefont
  {Born}}, \bibinfo {author} {\bibfnamefont {J.}~\bibnamefont {Schmitz}}, \
  and\ \bibinfo {author} {\bibfnamefont {M.}~\bibnamefont {Sperl}},\ }\href
  {\doibase 10.1038/s41526-017-0030-z} {\bibfield  {journal} {\bibinfo
  {journal} {NPJ Microgravity}\ }\textbf {\bibinfo {volume} {3}},\ \bibinfo
  {pages} {27} (\bibinfo {year} {2017})}\BibitemShut {NoStop}%
\bibitem [{\citenamefont {Harth}\ \emph {et~al.}(2018)\citenamefont {Harth},
  \citenamefont {Trittel}, \citenamefont {Wegner},\ and\ \citenamefont
  {Stannarius}}]{harth:18}%
  \BibitemOpen
  \bibfield  {author} {\bibinfo {author} {\bibfnamefont {K.}~\bibnamefont
  {Harth}}, \bibinfo {author} {\bibfnamefont {T.}~\bibnamefont {Trittel}},
  \bibinfo {author} {\bibfnamefont {S.}~\bibnamefont {Wegner}}, \ and\ \bibinfo
  {author} {\bibfnamefont {R.}~\bibnamefont {Stannarius}},\ }\href@noop {}
  {\bibfield  {journal} {\bibinfo  {journal} {Phys. Rev. Lett.}\ }\textbf
  {\bibinfo {volume} {120}},\ \bibinfo {pages} {214301} (\bibinfo {year}
  {2018})}\BibitemShut {NoStop}%
\bibitem [{\citenamefont {Hou}\ \emph {et~al.}(2008)\citenamefont {Hou},
  \citenamefont {Liu}, \citenamefont {Zhai}, \citenamefont {Sun}, \citenamefont
  {Lu}, \citenamefont {Garrabos},\ and\ \citenamefont {Evesque}}]{Hou2008}%
  \BibitemOpen
  \bibfield  {author} {\bibinfo {author} {\bibfnamefont {M.}~\bibnamefont
  {Hou}}, \bibinfo {author} {\bibfnamefont {R.}~\bibnamefont {Liu}}, \bibinfo
  {author} {\bibfnamefont {G.}~\bibnamefont {Zhai}}, \bibinfo {author}
  {\bibfnamefont {Z.}~\bibnamefont {Sun}}, \bibinfo {author} {\bibfnamefont
  {K.}~\bibnamefont {Lu}}, \bibinfo {author} {\bibfnamefont {Y.}~\bibnamefont
  {Garrabos}}, \ and\ \bibinfo {author} {\bibfnamefont {P.}~\bibnamefont
  {Evesque}},\ }\href@noop {} {\bibfield  {journal} {\bibinfo  {journal}
  {Microgravity Sci. and Tech.}\ }\textbf {\bibinfo {volume} {20}},\ \bibinfo
  {pages} {73} (\bibinfo {year} {2008})}\BibitemShut {NoStop}%
\bibitem [{\citenamefont {Kudrolli}\ and\ \citenamefont
  {Henry}(2000)}]{kudrolli:00}%
  \BibitemOpen
  \bibfield  {author} {\bibinfo {author} {\bibfnamefont {A.}~\bibnamefont
  {Kudrolli}}\ and\ \bibinfo {author} {\bibfnamefont {J.}~\bibnamefont
  {Henry}},\ }\href {\doibase 10.1103/PhysRevE.62.R1489} {\bibfield  {journal}
  {\bibinfo  {journal} {Phys. Rev. E}\ }\textbf {\bibinfo {volume} {62}},\
  \bibinfo {pages} {R1489} (\bibinfo {year} {2000})}\BibitemShut {NoStop}%
\bibitem [{\citenamefont {Rouyer}\ and\ \citenamefont
  {Menon}(2000)}]{rouyer:00}%
  \BibitemOpen
  \bibfield  {author} {\bibinfo {author} {\bibfnamefont {F.}~\bibnamefont
  {Rouyer}}\ and\ \bibinfo {author} {\bibfnamefont {N.}~\bibnamefont {Menon}},\
  }\href {\doibase 10.1103/PhysRevLett.85.3676} {\bibfield  {journal} {\bibinfo
   {journal} {Phys. Rev. Lett.}\ }\textbf {\bibinfo {volume} {85}},\ \bibinfo
  {pages} {3676} (\bibinfo {year} {2000})}\BibitemShut {NoStop}%
\bibitem [{\citenamefont {Blair}\ and\ \citenamefont
  {Kudrolli}(2003)}]{blair:03}%
  \BibitemOpen
  \bibfield  {author} {\bibinfo {author} {\bibfnamefont {D.~L.}\ \bibnamefont
  {Blair}}\ and\ \bibinfo {author} {\bibfnamefont {A.}~\bibnamefont
  {Kudrolli}},\ }\href {\doibase 10.1103/PhysRevE.67.041301} {\bibfield
  {journal} {\bibinfo  {journal} {Phys. Rev. E}\ }\textbf {\bibinfo {volume}
  {67}},\ \bibinfo {pages} {041301} (\bibinfo {year} {2003})}\BibitemShut
  {NoStop}%
\bibitem [{\citenamefont {Losert}\ \emph {et~al.}(1999)\citenamefont {Losert},
  \citenamefont {Cooper}, \citenamefont {Delour}, \citenamefont {Kudrolli},\
  and\ \citenamefont {Gollub}}]{losert:99}%
  \BibitemOpen
  \bibfield  {author} {\bibinfo {author} {\bibfnamefont {W.}~\bibnamefont
  {Losert}}, \bibinfo {author} {\bibfnamefont {D.~G.~W.}\ \bibnamefont
  {Cooper}}, \bibinfo {author} {\bibfnamefont {J.}~\bibnamefont {Delour}},
  \bibinfo {author} {\bibfnamefont {A.}~\bibnamefont {Kudrolli}}, \ and\
  \bibinfo {author} {\bibfnamefont {J.~P.}\ \bibnamefont {Gollub}},\ }\href
  {\doibase 10.1063/1.166442} {\bibfield  {journal} {\bibinfo  {journal}
  {Chaos}\ }\textbf {\bibinfo {volume} {9}},\ \bibinfo {pages} {682} (\bibinfo
  {year} {1999})}\BibitemShut {NoStop}%
\bibitem [{\citenamefont {Huan}\ \emph {et~al.}(2004)\citenamefont {Huan},
  \citenamefont {Yang}, \citenamefont {Candela}, \citenamefont {Mair},\ and\
  \citenamefont {Walsworth}}]{huan:04}%
  \BibitemOpen
  \bibfield  {author} {\bibinfo {author} {\bibfnamefont {C.}~\bibnamefont
  {Huan}}, \bibinfo {author} {\bibfnamefont {X.}~\bibnamefont {Yang}}, \bibinfo
  {author} {\bibfnamefont {D.}~\bibnamefont {Candela}}, \bibinfo {author}
  {\bibfnamefont {R.~W.}\ \bibnamefont {Mair}}, \ and\ \bibinfo {author}
  {\bibfnamefont {R.~L.}\ \bibnamefont {Walsworth}},\ }\href {\doibase
  10.1103/PhysRevE.69.041302} {\bibfield  {journal} {\bibinfo  {journal} {Phys.
  Rev. E}\ }\textbf {\bibinfo {volume} {69}},\ \bibinfo {pages} {041302}
  (\bibinfo {year} {2004})}\BibitemShut {NoStop}%
\bibitem [{\citenamefont {van Zon}\ \emph {et~al.}(2004)\citenamefont {van
  Zon}, \citenamefont {Kreft}, \citenamefont {Goldman}, \citenamefont
  {Miracle}, \citenamefont {Swift},\ and\ \citenamefont {Swinney}}]{vanZon:04}%
  \BibitemOpen
  \bibfield  {author} {\bibinfo {author} {\bibfnamefont {J.~S.}\ \bibnamefont
  {van Zon}}, \bibinfo {author} {\bibfnamefont {J.}~\bibnamefont {Kreft}},
  \bibinfo {author} {\bibfnamefont {D.~I.}\ \bibnamefont {Goldman}}, \bibinfo
  {author} {\bibfnamefont {D.}~\bibnamefont {Miracle}}, \bibinfo {author}
  {\bibfnamefont {J.~B.}\ \bibnamefont {Swift}}, \ and\ \bibinfo {author}
  {\bibfnamefont {H.~L.}\ \bibnamefont {Swinney}},\ }\href {\doibase
  10.1103/PhysRevE.70.040301} {\bibfield  {journal} {\bibinfo  {journal} {Phys.
  Rev. E}\ }\textbf {\bibinfo {volume} {70}},\ \bibinfo {pages} {040301}
  (\bibinfo {year} {2004})}\BibitemShut {NoStop}%
\bibitem [{\citenamefont {Scholz}\ and\ \citenamefont
  {P\"oschel}(2017)}]{scholz:17}%
  \BibitemOpen
  \bibfield  {author} {\bibinfo {author} {\bibfnamefont {C.}~\bibnamefont
  {Scholz}}\ and\ \bibinfo {author} {\bibfnamefont {T.}~\bibnamefont
  {P\"oschel}},\ }\href {\doibase 10.1103/PhysRevLett.118.198003} {\bibfield
  {journal} {\bibinfo  {journal} {Phys. Rev. Lett.}\ }\textbf {\bibinfo
  {volume} {118}},\ \bibinfo {pages} {198003} (\bibinfo {year}
  {2017})}\BibitemShut {NoStop}%
\bibitem [{\citenamefont {Luding}\ \emph {et~al.}(1998)\citenamefont {Luding},
  \citenamefont {Huthmann}, \citenamefont {McNamara},\ and\ \citenamefont
  {Zippelius}}]{luding:98}%
  \BibitemOpen
  \bibfield  {author} {\bibinfo {author} {\bibfnamefont {S.}~\bibnamefont
  {Luding}}, \bibinfo {author} {\bibfnamefont {M.}~\bibnamefont {Huthmann}},
  \bibinfo {author} {\bibfnamefont {S.}~\bibnamefont {McNamara}}, \ and\
  \bibinfo {author} {\bibfnamefont {A.}~\bibnamefont {Zippelius}},\ }\href
  {\doibase 10.1103/PhysRevE.58.3416} {\bibfield  {journal} {\bibinfo
  {journal} {Phys. Rev. E}\ }\textbf {\bibinfo {volume} {58}},\ \bibinfo
  {pages} {3416} (\bibinfo {year} {1998})}\BibitemShut {NoStop}%
\bibitem [{\citenamefont {Kanzaki}\ \emph {et~al.}(2010)\citenamefont
  {Kanzaki}, \citenamefont {Hidalgo}, \citenamefont {Maza},\ and\ \citenamefont
  {Pagonabarraga}}]{kanzaki:10}%
  \BibitemOpen
  \bibfield  {author} {\bibinfo {author} {\bibfnamefont {T.}~\bibnamefont
  {Kanzaki}}, \bibinfo {author} {\bibfnamefont {R.~C.}\ \bibnamefont
  {Hidalgo}}, \bibinfo {author} {\bibfnamefont {D.}~\bibnamefont {Maza}}, \
  and\ \bibinfo {author} {\bibfnamefont {I.}~\bibnamefont {Pagonabarraga}},\
  }\href {\doibase 10.1088/1742-5468/2010/06/P06020} {\bibfield  {journal}
  {\bibinfo  {journal} {J. Stat. Mech.: Theory Exp.}\ }\textbf {\bibinfo
  {volume} {2010}},\ \bibinfo {pages} {P06020} (\bibinfo {year}
  {2010})}\BibitemShut {NoStop}%
\bibitem [{\citenamefont {Bodrova}\ \emph {et~al.}(2015)\citenamefont
  {Bodrova}, \citenamefont {Chechkin}, \citenamefont {Cherstvy},\ and\
  \citenamefont {Metzler}}]{bodrova:15}%
  \BibitemOpen
  \bibfield  {author} {\bibinfo {author} {\bibfnamefont {A.}~\bibnamefont
  {Bodrova}}, \bibinfo {author} {\bibfnamefont {A.~V.}\ \bibnamefont
  {Chechkin}}, \bibinfo {author} {\bibfnamefont {A.~G.}\ \bibnamefont
  {Cherstvy}}, \ and\ \bibinfo {author} {\bibfnamefont {R.}~\bibnamefont
  {Metzler}},\ }\href {\doibase 10.1039/C5CP02824H} {\bibfield  {journal}
  {\bibinfo  {journal} {Phys. Chem. Chem. Phys.}\ }\textbf {\bibinfo {volume}
  {17}},\ \bibinfo {pages} {21791} (\bibinfo {year} {2015})}\BibitemShut
  {NoStop}%
\bibitem [{\citenamefont {Brey}\ and\ \citenamefont
  {Ruiz-Montero}(1999)}]{brey:99}%
  \BibitemOpen
  \bibfield  {author} {\bibinfo {author} {\bibfnamefont {J.~J.}\ \bibnamefont
  {Brey}}\ and\ \bibinfo {author} {\bibfnamefont {M.~J.}\ \bibnamefont
  {Ruiz-Montero}},\ }\href {\doibase 10.1016/S0010-4655(99)00331-8} {\bibfield
  {journal} {\bibinfo  {journal} {Comput. Phys. Commun.}\ }\textbf {\bibinfo
  {volume} {121-122}},\ \bibinfo {pages} {278} (\bibinfo {year}
  {1999})}\BibitemShut {NoStop}%
\bibitem [{\citenamefont {Brey}\ \emph {et~al.}(1999)\citenamefont {Brey},
  \citenamefont {Cubero},\ and\ \citenamefont {Ruiz-Montero}}]{brey:99-2}%
  \BibitemOpen
  \bibfield  {author} {\bibinfo {author} {\bibfnamefont {J.~J.}\ \bibnamefont
  {Brey}}, \bibinfo {author} {\bibfnamefont {D.}~\bibnamefont {Cubero}}, \ and\
  \bibinfo {author} {\bibfnamefont {M.~J.}\ \bibnamefont {Ruiz-Montero}},\
  }\href {\doibase 10.1103/PhysRevE.59.1256} {\bibfield  {journal} {\bibinfo
  {journal} {Phys. Rev. E}\ }\textbf {\bibinfo {volume} {59}},\ \bibinfo
  {pages} {1256} (\bibinfo {year} {1999})}\BibitemShut {NoStop}%
\bibitem [{SUP()}]{SUPPvideo}%
  \BibitemOpen
  \href {http://link.aps.org/supplemental/00.0000/PhysRevLett.000.000001}
  {}\bibinfo {howpublished} {See the supplementary video at
  \url{http://link.aps.org/supplemental/00.0000/PhysRevLett.000.000001} for
  particle dynamics several seconds before and after the start of the
  cooling.}\BibitemShut {Stop}%
\bibitem [{\citenamefont {Carnahan}\ and\ \citenamefont
  {Starling}(1969)}]{Carnahan1969}%
  \BibitemOpen
  \bibfield  {author} {\bibinfo {author} {\bibfnamefont {N.~F.}\ \bibnamefont
  {Carnahan}}\ and\ \bibinfo {author} {\bibfnamefont {K.~E.}\ \bibnamefont
  {Starling}},\ }\href@noop {} {\bibfield  {journal} {\bibinfo  {journal} {J.
  Chem. Phys.}\ }\textbf {\bibinfo {volume} {51}},\ \bibinfo {pages} {635}
  (\bibinfo {year} {1969})}\BibitemShut {NoStop}%
\bibitem [{\citenamefont {Santos}\ \emph {et~al.}(2011)\citenamefont {Santos},
  \citenamefont {Kremer},\ and\ \citenamefont {dos Santos}}]{santos:11}%
  \BibitemOpen
  \bibfield  {author} {\bibinfo {author} {\bibfnamefont {A.}~\bibnamefont
  {Santos}}, \bibinfo {author} {\bibfnamefont {G.~M.}\ \bibnamefont {Kremer}},
  \ and\ \bibinfo {author} {\bibfnamefont {M.}~\bibnamefont {dos Santos}},\
  }\href {\doibase 10.1063/1.3558876} {\bibfield  {journal} {\bibinfo
  {journal} {Phys. Fluids}\ }\textbf {\bibinfo {volume} {23}},\ \bibinfo
  {pages} {030604} (\bibinfo {year} {2011})}\BibitemShut {NoStop}%
\bibitem [{\citenamefont {Bodrova}\ and\ \citenamefont
  {Brilliantov}(2012)}]{bodrova:12}%
  \BibitemOpen
  \bibfield  {author} {\bibinfo {author} {\bibfnamefont {A.}~\bibnamefont
  {Bodrova}}\ and\ \bibinfo {author} {\bibfnamefont {N.}~\bibnamefont
  {Brilliantov}},\ }\href {\doibase 10.1007/s10035-012-0319-2} {\bibfield
  {journal} {\bibinfo  {journal} {Granul. Matter}\ }\textbf {\bibinfo {volume}
  {14}},\ \bibinfo {pages} {85} (\bibinfo {year} {2012})}\BibitemShut {NoStop}%
\bibitem [{\citenamefont {Ouellette}\ \emph {et~al.}(2006)\citenamefont
  {Ouellette}, \citenamefont {Xu},\ and\ \citenamefont
  {Bodenschatz}}]{ouellette:06}%
  \BibitemOpen
  \bibfield  {author} {\bibinfo {author} {\bibfnamefont {N.~T.}\ \bibnamefont
  {Ouellette}}, \bibinfo {author} {\bibfnamefont {H.}~\bibnamefont {Xu}}, \
  and\ \bibinfo {author} {\bibfnamefont {E.}~\bibnamefont {Bodenschatz}},\
  }\href {\doibase 10.1007/s00348-005-0068-7} {\bibfield  {journal} {\bibinfo
  {journal} {Exp. Fluids}\ }\textbf {\bibinfo {volume} {40}},\ \bibinfo {pages}
  {301} (\bibinfo {year} {2006})}\BibitemShut {NoStop}%
\bibitem [{\citenamefont {Jaqaman}\ \emph {et~al.}(2008)\citenamefont
  {Jaqaman}, \citenamefont {Loerke}, \citenamefont {Mettlen}, \citenamefont
  {Kuwata}, \citenamefont {Grinstein}, \citenamefont {Schmid},\ and\
  \citenamefont {Danuser}}]{jaqaman:08}%
  \BibitemOpen
  \bibfield  {author} {\bibinfo {author} {\bibfnamefont {K.}~\bibnamefont
  {Jaqaman}}, \bibinfo {author} {\bibfnamefont {D.}~\bibnamefont {Loerke}},
  \bibinfo {author} {\bibfnamefont {M.}~\bibnamefont {Mettlen}}, \bibinfo
  {author} {\bibfnamefont {H.}~\bibnamefont {Kuwata}}, \bibinfo {author}
  {\bibfnamefont {S.}~\bibnamefont {Grinstein}}, \bibinfo {author}
  {\bibfnamefont {S.~L.}\ \bibnamefont {Schmid}}, \ and\ \bibinfo {author}
  {\bibfnamefont {G.}~\bibnamefont {Danuser}},\ }\href {\doibase
  10.1038/nmeth.1237} {\bibfield  {journal} {\bibinfo  {journal} {Nat.
  Methods}\ }\textbf {\bibinfo {volume} {5}},\ \bibinfo {pages} {695} (\bibinfo
  {year} {2008})}\BibitemShut {NoStop}%
\bibitem [{\citenamefont {Besseling}\ \emph {et~al.}(2009)\citenamefont
  {Besseling}, \citenamefont {Isa}, \citenamefont {Weeks},\ and\ \citenamefont
  {Poon}}]{besseling:09}%
  \BibitemOpen
  \bibfield  {author} {\bibinfo {author} {\bibfnamefont {R.}~\bibnamefont
  {Besseling}}, \bibinfo {author} {\bibfnamefont {L.}~\bibnamefont {Isa}},
  \bibinfo {author} {\bibfnamefont {E.~R.}\ \bibnamefont {Weeks}}, \ and\
  \bibinfo {author} {\bibfnamefont {W.~C.}\ \bibnamefont {Poon}},\ }\href
  {\doibase 10.1016/j.cis.2008.09.008} {\bibfield  {journal} {\bibinfo
  {journal} {Adv. Colloid Interface Sci.}\ }\textbf {\bibinfo {volume} {146}},\
  \bibinfo {pages} {1} (\bibinfo {year} {2009})}\BibitemShut {NoStop}%
\bibitem [{\citenamefont {Dalitz}\ \emph {et~al.}(2017)\citenamefont {Dalitz},
  \citenamefont {Schramke},\ and\ \citenamefont {Jeltsch}}]{dalitz:17}%
  \BibitemOpen
  \bibfield  {author} {\bibinfo {author} {\bibfnamefont {C.}~\bibnamefont
  {Dalitz}}, \bibinfo {author} {\bibfnamefont {T.}~\bibnamefont {Schramke}}, \
  and\ \bibinfo {author} {\bibfnamefont {M.}~\bibnamefont {Jeltsch}},\ }\href
  {\doibase 10.5201/ipol.2017.208} {\bibfield  {journal} {\bibinfo  {journal}
  {Image Processing On Line}\ }\textbf {\bibinfo {volume} {7}},\ \bibinfo
  {pages} {184} (\bibinfo {year} {2017})}\BibitemShut {NoStop}%
\end{thebibliography}%
\end{document}